\documentclass[a4paper,fleqn,usenatbib]{mnras}

\usepackage{mathptmx}

\usepackage[T1]{fontenc}
\usepackage{ae,aecompl}

\usepackage{graphicx}	
\usepackage{amsmath}	
\usepackage{amssymb}	

\newcommand{\feh}{$\rm[Fe/H]\;$}
\newcommand{\mfeh}{$\langle \rm[Fe/H]\rangle\;$}
\newcommand{\ews}{OWS}

\newcommand{\revise}[1]{\textcolor{black}{#1}} 

\title[Morphology of the GSS progenitor]{Formation of the Andromeda Giant Stream: Asymmetric Structure and Disc Progenitor}

\author[T. Kirihara et al.]{
T. Kirihara,$^{1}$\thanks{E-mail: kirihara@ccs.tsukuba.ac.jp}
Y. Miki,$^{2,3}$
M. Mori,$^{1,2}$
\revise{T. Kawaguchi,$^{4}$
and R. M. Rich$^{5}$}
\\
$^{1}$Faculty of Pure and Applied Physics, University of Tsukuba, Tennodai 1-1-1, Tsukuba, Ibaraki, Japan\\
$^{2}$Center for Computational Sciences, University of Tsukuba, Tennodai 1-1-1, Tsukuba, Ibaraki, Japan\\
$^{3}$CREST, JST, Tennodai 1-1-1, Tsukuba, Ibaraki, Japan\\
$^{4}$Sapporo Medical University, S1W17, Chuo-ku, Sapporo 060-8556, Hokkaido, Japan\\
\revise{$^{5}$Department of Physics and Astronomy, University of California, Los Angeles, CA 90095, USA}
}

\date{Accepted XXX. Received YYY; in original form ZZZ}

\pubyear{2016}

\begin{document}
\label{firstpage}
\pagerange{\pageref{firstpage}--\pageref{lastpage}}
\maketitle

\begin{abstract}

We focus on the evidence of a past minor merger discovered in the halo of the Andromeda galaxy (M31). 
Previous $N$--body studies have \revise{enjoyed moderate success in producing the} observed giant stellar stream (GSS) and stellar shells in M31\revise{'s halo}. 
The observed distribution of stars in the halo of M31 shows an asymmetric surface brightness profile across the GSS; however, \revise{the effect of the morphology of the progenitor galaxy on the internal structure of the GSS requires further investigation in theoretical studies. }
To investigate the physical connection between the characteristic surface brightness in the GSS and the morphology of the progenitor dwarf galaxy, we \revise{systematically vary the thickness, rotation velocity and initial inclination of the disc dwarf galaxy in $N$--body simulations. }
\revise{The} formation of the observed structures \revise{appears to be dominated by} the progenitor's rotation. 
\revise{Besides reproducing the observed GSS and two shells in detail, we predict additional structures for further observations. }
We predict the detectability of the progenitor's stellar core in \revise{the} phase-space density distribution, azimuthal metallicity gradient of the western shell-like structure and an additional extended shell \revise{in} the north-western direction that \revise{may constrain} the properties of the progenitor galaxy. 

\end{abstract}



\section{Introduction}

\revise{The} Local \revise{Group} is \revise{a natural laboratory for investigating} the formation and evolution of galaxies \revise{and comparing the observations with theoretical studies}. 
\revise{According to the generally accepted cold dark matter model, a snapshot} of the Local \revise{Group} should \revise{record} a history of the hierarchical \revise{structural} formation of the universe. 
In fact, \revise{by studying the} spatial, kinematic and metallicity \revise{distributions} of sub-structures such as tidal debris and dwarf satellite galaxies in \revise{their} host halo\revise{, we can probe the} formation history of galaxies, \revise{the} density profile of the host galaxy and accretion history of massive black holes (MBHs) associated with satellite galaxies. 

Recent deep photometric observations of the halo of the Andromeda galaxy (M31) have discovered a wealth of faint structures\revise{,} including past and on-going galaxy mergers \citep{Ibata2001,Irwin2005,McConnachie2009,Martin2013}. 
A giant stellar stream (GSS) and fan-like stellar structures have been discovered in the halo of M31. 
Its spatial, metallicity and line-of-sight velocity distribution have been observed in detail \citep{Ibata2004,Ferguson2004,Kalirai2006,Gilbert2009}. 
\revise{Wider and deeper surveys of M31's halo region those of} PAndAS (Pan--Andromeda Archaeological Survey) \citep{McConnachie2009,Martin2013} and \revise{the} SPLASH (Spectroscopic and Photometric Landscape of Andromeda's Stellar Halo) survey \citep{Kalirai2010,Tollerud2012}. 
In particular, the GSS \revise{lies south-east of M31's centre} and extends more than $120$~kpc along the line-of-sight \citep{McConnachie2003,Conn2016}. 
\revise{The} fan-like structures spread \revise{to the north-east and west} side of M31 and \revise{with approximate radii of} $30$~kpc \citep{Fardal2007}. 
Various works have explored the formation of these structures \revise{by colliding two galaxies in} $N$--body simulations\revise{,} assuming minor \citep{Fardal2007} or major merger scenario \citep{Hammer2010}. 
We adopt a past radial interaction model of a dwarf satellite galaxy\revise{, which well reproduces almost all} of these structures \citep{Fardal2007,Mori2008,Fardal2013,Sadoun2014,Miki2014,Kirihara2014,Miki2016}. 

The total mass of the progenitor has been estimated \revise{by} several approaches. 
The lower limit of its stellar mass\revise{, estimated from the} kinematics and luminosity of the GSS\revise{, is $10^8M_{\sun}$} \citep{Font2006}. 
\citet{Mori2008} reported the upper limit of its dynamical mass \revise{as} $5\times 10^9M_{\sun}$\revise{. They considered the effect of} dynamical friction on the thickness of M31\revise{'s disc}. 
\citet{Fardal2013} and \citet{Miki2016} also examined the stellar mass of the satellite progenitor assuming a spherical progenitor galaxy\revise{. Their best-fitting value was approximately} $3-4\times10^9M_{\sun}$. 

\begin{figure}
  \includegraphics[width=\columnwidth]{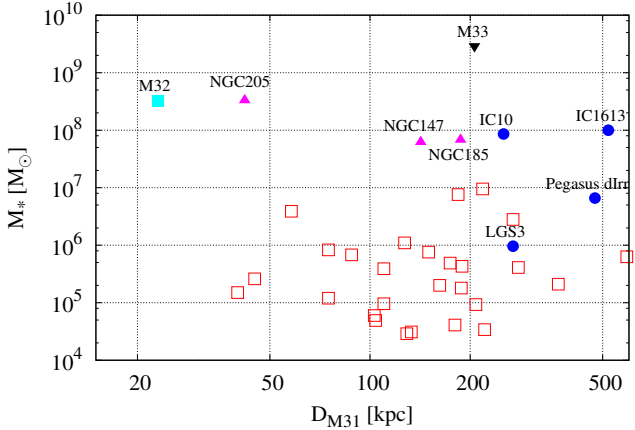}
  \caption{Stellar mass $M_*$ and morphological \revise{types} of the observed dwarf galaxies surrounding M31 as \revise{functions of} projected distance from the \revise{centre} of M31 $\rm{D}_{\rm M31}$(data are taken from \citet{McConnachie2012}). 
\revise{Symbols indicate} dSphs (open red squares), dIrrs (filled blue circles), dEs (filled magenta triangles), \revise{a} compact elliptical \revise{galaxy} (filled cyan square) and \revise{an} Sb galaxy (inverted filled triangle). 
\label{figure1}
  }
\end{figure}

Fig.~\ref{figure1} shows the stellar mass of the observed dwarf galaxies around M31 \revise{versus} the projected distance from the \revise{centre} of M31. 
\revise{The predicted} mass range of the progenitor dwarf galaxy \revise{is} dominated by dwarf \revise{ellipticals} and dwarf \revise{irregulars}. 
On the contrary, all such satellite galaxies in M31 \revise{have} rotating stellar and/or gas \revise{components} \citep{McConnachie2012}. 
Even M32\revise{, which is} classified \revise{as a} compact elliptical\revise{,} has a rotating stellar component with \revise{a} measured velocity of $55$ km s$^{-1}$ \citep{Bender1996}. 
\revise{M33 is a disc galaxy with no clear bulge. 
Disc galaxies comprise a large proportion of the less massive galaxies in the local universe \citep{Moffett2016}. 
}
Although previous studies have usually \revise{assumed} a spherical, non-rotating progenitor galaxy, these reasonable conditions motivate us to assume a disc-like galaxy as the GSS progenitor. 

Using star count maps around M31, \citet{McConnachie2003} \revise{analysed the} surface brightness profile \revise{in the direction orthogonal to major axis} of the GSS\revise{. They obtained} an asymmetric surface brightness of the GSS\revise{, with sharply decreased star counts at the north-eastern side of the GSS (viewed from the most luminous direction of the GSS).}
On the other hand, \revise{the surface brightness distribution extends widely and smoothly at the western side}. 
This asymmetrical structure of the GSS has \revise{never emerged in} simulations \revise{that assume the} infall of a spherical, non-rotating dwarf galaxy (e.g. Fig.~\ref{histall}d\revise{; \cite{Gilbert2007}}). 
The characteristic surface brightness profile of the GSS would be an excellent tracer of the morphology of the disrupted progenitor galaxy. 
\revise{To examine the disc merger scenario and identify the progenitor conditions that would lead to complicated evolution, we must systematically scrutinize a large parameter space. }
\revise{Few disc satellite models have generated an asymmetric structure for the GSS, moreover, they have not reproduced the observed shape \citep{Fardal2008,Sadoun2014}}. 
For example, \citet{Fardal2008} \revise{reported} two arc-like structures on the eastern side of the GSS (\revise{they resemble streams} C and D in \citet{Ibata2007})\revise{. Notably, the} sharp edge-like structure \revise{at the eastern side of the} GSS \revise{has not been} reproduced. 

\revise{Owing to its estimated mass,} the progenitor galaxy is \revise{unlikely to be a} nearby dSph\revise{, which has} mass-to-luminosity \revise{ratios} of $\sim$ 100. 
\revise{Initially, the} progenitor \revise{is thought to have inhabited} a dark matter halo, with \revise{a} mass-to-luminosity ratio ($M/L=2$--$5$). \revise{Similar $M/L$ values have been reported for} local satellite dwarf galaxies (see fig.~11 of \citet{McConnachie2012}). 
On the other hand, \revise{most previous works have ignored the} dark matter halo component of the progenitor galaxy\revise{, because this component was rationalized to have} been stripped before the first interaction with M31. 
However, \revise{if an inner} region of the dark matter halo \revise{was gravitationally strongly bound, it} would survive the collision \revise{with} M31. 
\revise{Further it is not understood how the collision would disperse the dark matter halo component through the host halo. }

\revise{The formation history of galaxies can be inferred from the spatial} distribution of \revise{heavy elements in their stars. }
\citet{Ibata2007} \revise{obtained} the metallicity distribution \revise{in the southern area of M31 from colour}-magnitude diagrams\revise{. They} suggested a clear metallicity \revise{difference} between \revise{an} eastern high-surface-brightness region (metal-rich) and \revise{a} faint western region (metal-poor) of the GSS. 
\revise{Similar trends in the GSS appear in spectroscopic} measurements of the metallicity distribution based on \revise{the} \ion{Ca}{II} triplet absorption lines \citep{Gilbert2009}. 
In addition, \citet{Fardal2012} measured the metallicity distribution of the western shell along the minor axis of M31's disc. 
Only recently, \citet{Conn2016} observed \revise{the} radial metallicity distribution along the GSS\revise{. }
\revise{Radial metallicity gradients are commonly observed in dwarf galaxies} \citep{Koleva2009}
\revise{, and are also known in nearby disc galaxies \citep{Magrini2016}.}
\revise{The present-day metallicity gradient of the GSS could conceivably have originated} in the progenitor dwarf galaxy \citep{Fardal2008,Miki2016}. 

\revise{In surveys of $N$--body simulations, the present paper explores galaxy collisions} between M31 and a dwarf satellite galaxy composed of a stellar disc, a stellar bulge and a dark matter halo component\revise{. The aim is} to reproduce the asymmetric surface brightness of the GSS. 
In \S \ref{sec:observation}, we describe the observational data and \revise{their} treatment, including our simple analysis of the asymmetric surface brightness profile of the GSS. 
In \S \ref{sec:simulation}, we introduce our modelling of the M31 \revise{potential}, \revise{the} $N$--body satellite progenitor and the numerical model for systematic surveys. 
The results of numerical simulations and quantitative \revise{comparisons} with observed data are displayed in \S \ref{sec:NR}. 
\revise{The} metallicity distribution, distribution of \revise{the} disrupted dark matter halo component of the progenitor, the position of \revise{a hypothetical} MBH \revise{(initially centralised in the progenitor galaxy)} and \revise{the} extended stellar shell at the north-western area of M31 are described in \S \ref{sec:discussions}\revise{. We summarise} our findings in \S \ref{sec:summary}.

\section{Observed Structures}
\label{sec:observation}
\subsection{Spatial faint structures around M31}

Merger remnants \revise{can reveal the properties (mass and morphological type) not only} of the host galaxy\revise{,} but also \revise{of} the disrupted progenitor galaxy. 
Some of the faint stellar structures in \revise{M31's halo} have been well reproduced \revise{by} an on-going merger of a satellite galaxy \citep{Fardal2007,Mori2008,Fardal2008,Fardal2013,Miki2014,Sadoun2014,Kirihara2014,Miki2016}. 
\citet{Irwin2005} \revise{intensively} surveyed the M31 halo \revise{with} the Isaac Newton Telescope Wide--Field Camera (INT/WFC). 
Their map covers \revise{a 4$\degr$} elliptical region of the semi-major axis with \revise{an} aspect ratio of 5:3 and an additional \revise{$\sim$10 [degree$^2$]} extension towards the south of M31. 
We \revise{analyse their stellar dataset} \revise{in comparisons with our} numerical simulations. 

The star count maps include non-GSS components such as \revise{the} halo stars of M31, foreground halo stars of MW, and other substructures around M31. 
\revise{To clarify the GSS structure, we} simply subtract a constant background as follows. 
\revise{We first calculate the GSS background by removing} the area of M31's disc, the eastern and the western shell, the GSS and several stellar substructures from the star count map. 
Next, we \revise{average the} star counts \revise{per} cell\revise{, excluding the above non-background area}. 
The \revise{stellar} background \revise{count} is $1.18$ \revise{stars} per cell. 
\revise{This count is subtracted from the star counts in each cell. }
\revise{This treatment provides a} clearer structure of the GSS.

\subsection{Asymmetric structure of the GSS}
\label{sec:asymmetry}

\revise{In their analysis of the star count maps,} \citet{McConnachie2003} \revise{found an asymmetric surface brightness across the GSS}. 
\revise{Specifically, they reported a sharp increase in star counts at} the eastern side of the GSS\revise{, relative to} the western side. 
\revise{To compare our} simulated GSS with the observed GSS, we \revise{reanalyse} the observed data and obtain \revise{the} azimuthal surface brightness profile of the GSS. 

\begin{figure}
  \includegraphics[width=\columnwidth]{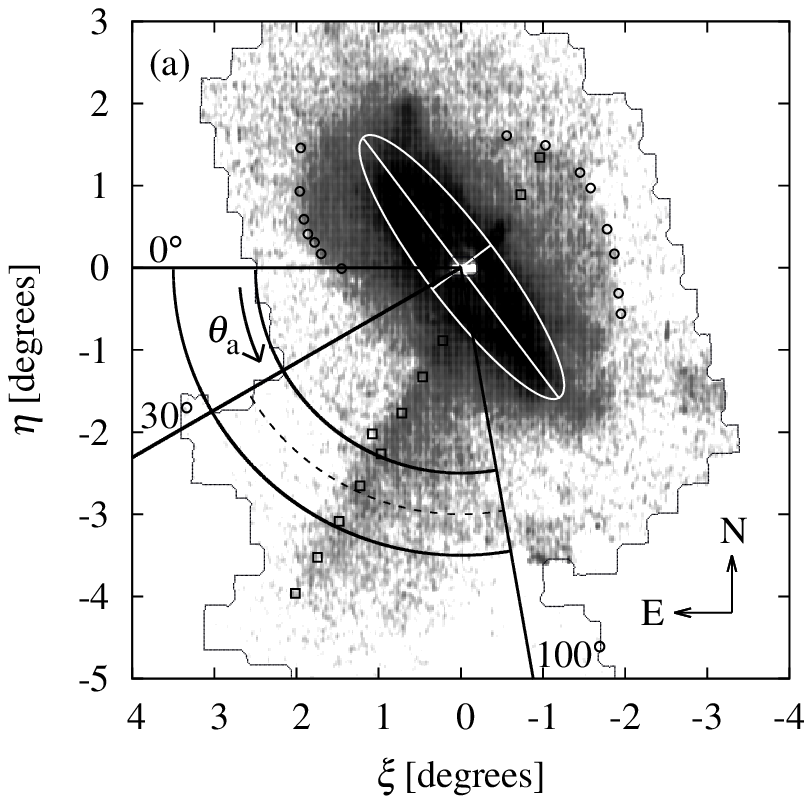}
  \includegraphics[width=\columnwidth]{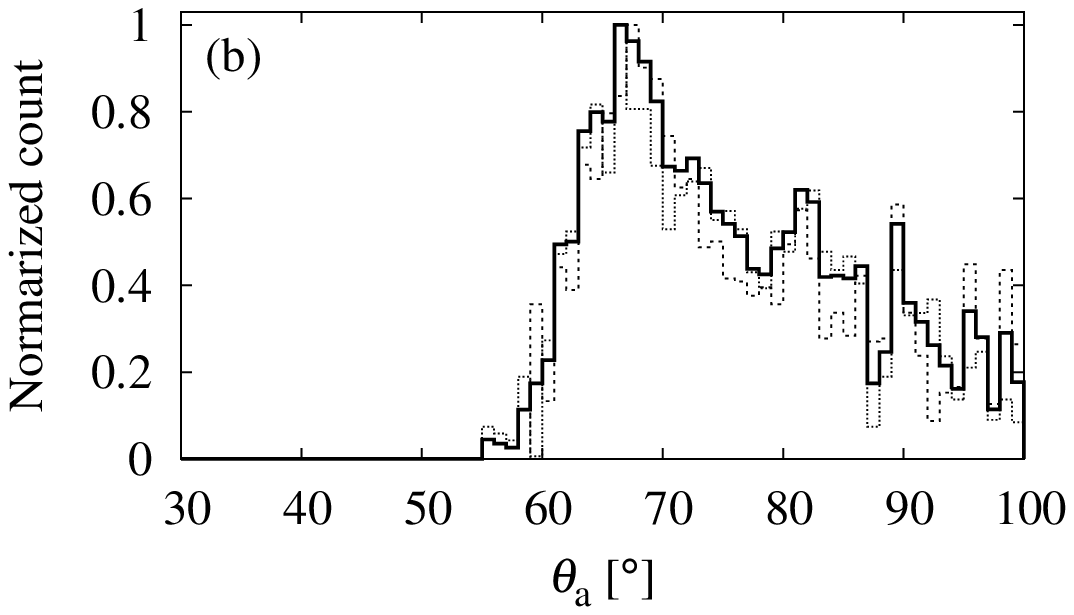}
  \caption{(a) Background-subtracted stellar count map of the halo of M31 \citep{Irwin2005}. 
    \revise{Arcs denote} the observed edges of the eastern and western shells and squares \revise{locate in} the previously observed area of the GSS \citep{McConnachie2003,Guhathakurta2006,Fardal2007}. 
    The white ellipse \revise{traces the disc of M31}. 
    The azimuthal angle $\theta_a$ is taken from the eastern direction \revise{in M31-centred coordinates}. 
    (b) Azimuthal star \revise{count} distribution of the GSS. 
    The \revise{analysis area is a} portion of an annulus ($30\degr<\theta_a<100\degr$)\revise{, as} shown in panel (a). 
    The dotted and dashed lines \revise{profile} the inner and outer \revise{regions} of the annulus, respectively. 
    \label{figure2}
  }
\end{figure}

\revise{Within the} radius $R=2.5$$\revise{\degr}$ ($\sim$30~kpc)\revise{, the pure GSS component is obscured by} superposition of M31's disc and the clumpy stellar structures. 
\revise{Beyond $R=3.5\degr$, the star counts are insufficient for a proper analysis. }
Therefore, \revise{we analyse the region  $2.5\degr < R <3.5\degr$ from the centre of M31 over the azimuthal angular range $30\degr<\theta_a<100\degr$ (east to south).} 
\revise{The radial distance is divided into inner and outer areas separated at $R = 3.0\degr$. 
The analysed area is enclosed by solid lines in Fig.~\ref{figure2}a. }

Fig.~\ref{figure2}b shows the azimuthal \revise{distribution of} star counts \revise{in} the GSS. 
\revise{Hereafter, we define the} brightest azimuthal angle \revise{($\theta_a\sim65\degr$) as the GSS axis}. 
\revise{The surface brightness gradually decreases at the} western side (\revise{larger} $\theta_a$) \revise{of} the GSS axis\revise{, but increases sharply at} the eastern side. 
\revise{Each side of this} asymmetric star count profile \revise{is fitted} by an asymmetrical exponential function \revise{proportional} to exp$(-|d \theta_a|/\delta_{\rm{obs}}^{\pm})$\revise{, where} $\delta_{\rm{obs}}^{-}$ and $\delta_{\rm{obs}}^{+}$ are the observed \revise{widths} of the eastern edge and the broad western structure, respectively. 
The fitted values are \revise{summarised} in \autoref{tab:table2}. 
The significant \revise{width} difference \revise{between the two sides confirms the strongly asymmetric spatial distribution. }

\begin{table}
 \centering
 \begin{minipage}{84mm}
  \caption{Fitted values of the GSS surface brightness \label{tab:table2}}
  \begin{tabular}{@{}cccc@{}}
  \hline
  Region & GSS axis & Eastern Edge& Western Broad\\
         &  $\theta_{\rm axis}$ [$\degr$]          & width $\delta_{\rm obs}^-$ [$\degr$] & width $\delta_{\rm obs}^+$ [$\degr$]\\
 \hline
Inner & 64.0 & 2.74 & 24.1\\
Outer & 65.1 & 3.34 & 21.9\\
\hline
\end{tabular}
\end{minipage}
\end{table}

\section{Numerical models}
\label{sec:simulation}
\subsection{M31 potential model}
\label{sec:M31model}

In this work, we assume a fixed gravitational potential model for M31 with a Hernquist bulge \citep{Hernquist1990}, an exponential disc and a dark matter halo with \revise{an} NFW profile \revise{\citep{Navarro1996}}. 
The scale radius and total mass of the bulge component are $0.61$~kpc and $3.24 \times 10^{10}M_{\sun}$, respectively. 
The \revise{M31} disc \revise{is assigned a} scale height of $0.6$~kpc, \revise{a} radial scale length of $5.4$~kpc and \revise{a} total mass of $3.66 \times 10^{10}M_{\sun}$. 
The inclination and position \revise{angle} of the \revise{M31} disc are $77\degr$ and $37\degr$, respectively \citep{Geehan2006}. 
The \revise{NFW halo has a} scale radius of $7.63$~kpc, and \revise{a} scale density \revise{of} $6.17 \times 10^7 M_{\sun} \rm{kpc}^{-3}$. 
\revise{Under these} conditions\revise{,} the M31 model nicely reproduces the observed surface brightness of the bulge and disc components, the velocity dispersion of the bulge and the rotation curve of the disc \citep{Geehan2006,Fardal2007}. 

\revise{Also using full $N$--body simulations,} \citet{Mori2008} examined the dynamic response of the interaction \revise{between M31 and a less massive progenitor} ($\lesssim5\times 10^9 M_{\sun}$)\revise{. They found} little change in the gravitational potential of M31. 
Therefore, M31 \revise{can be feasibly treated} as a fixed gravitational potential, and \revise{the} above-mentioned \revise{M31} parameters \revise{have been adopted in} previous studies using the same initial orbital variables of the progenitor \citep{Fardal2007,Fardal2008,Miki2016}. 

\subsection{$N$-body satellite models}
\label{sec:satellite_models}

\revise{To} elucidate the origin of the asymmetric \revise{GSS} surface brightness, we \revise{collide} M31 with a disc dwarf satellite galaxy\revise{, which is} a self-consistent $N$--body \revise{disc with stars and dark matter}. 
The initial position and velocity \revise{vectors} of the progenitor galaxy are taken from \citet{Fardal2007}. 

An equilibrium model \revise{of the} disc dwarf galaxy \revise{is} constructed \revise{using in} the public code GalactICS \citep{Kuijken1995}, which generates a self-consistent bulge--disc--halo $N$--body system. 
\revise{As a tentative} disc progenitor model for \revise{the THIN model, we adopt the downsized model of the M31 Model A presented in \citet{Widrow2003}, which reduces the M31 mass by a factor of 100. }
\revise{This model treats} the density profile of the dark matter halo \revise{as} a lowered Evans profile \citep{Kuijken1994}\revise{, which satisfies} an isothermal distribution function \revise{with a characteristic radius $r_a$}. 
The disc component follows \revise{an} exponential density profile \revise{with a scale length $R_{\rm d}$ in the radial direction, and an} isothermal profile \revise{with a scale height $Z_{\rm d}$ in the vertical direction}. 
The bulge component is \revise{modelled as} a King sphere. 

\revise{We now summarise} the input parameter set of the disc dwarf progenitor models. 
\revise{In} GalactICS, the length, velocity and mass \revise{units} are 1~kpc, $100$ km s$^{-1}$ and $2.325\times 10^{9}M_{\sun}$, respectively (see \cite{Kuijken1995} and \cite{Widrow2003}). 
\revise{The halo} parameters \revise{input to} THIN are the central potential $\Phi(0)=-1.483$, the central velocity dispersion $V_0=0.952$ and $r_a=0.981$. 
The total mass\revise{, scale length and scale height} of the disc component \revise{are $M_{\rm{d}}=0.318$, $R_{\rm{d}}=1.11$ and $Z_{\rm{d}}=0.13$, respectively. }
The outermost cutoff radius of the disc $R_{\rm{outer}}$ is 8.0, \revise{with a} truncation length \revise{of} $0.2$. 
The \revise{parameters of the} bulge component \revise{are the central density ($\rho_0=6.68$), velocity dispersion ($\sigma_b=0.508$) and cutoff potential ($\psi_{\rm{cut}}=-0.835$). }
\revise{The cutoff potential controls} the extent of the bulge.

\begin{table*}
  \caption{Properties of the progenitor models\label{tab:table4}}
  \begin{tabular}{@{}cccccccccccc@{}}
  \hline
Model & THIN & THICK  & THICK2 & THICK3 & THICK4 & THICK5 & THICK6 & THICK7 & THICK8 & THICK9 & HOT \\
\hline
$Z_{\rm{d}}$& 0.13 & 0.52 & 0.52 & 0.52 & 0.52 & 0.52 & 0.52 & 0.52 & 0.52 & 0.52 & 1.11 \\
$\Psi_{0}$  &-1.483&-1.483&-1.400&-1.450&-1.500&-1.550&-1.600&-1.650&-1.700&-1.750&-1.483 \\
$V_0$       & 0.952& 0.952& 1.000& 1.100& 1.200& 1.300& 1.400& 1.500& 1.600& 1.650& 0.952 \\
\hline
\end{tabular}
\end{table*}

\revise{As is well known, the} nearby dwarf galaxies \revise{are relatively thick} \citep{Spolaor2010,Toloba2011}. 
Accordingly, we construct two additional models \revise{(THICK and HOT), in which the} scale \revise{heights are $Z_{\rm{d}}=0.52$ ($4$ times the scale height of the disc of THIN) and $Z_{\rm{d}}=1.11$ (the same length as the scale length), respectively}. 
The input \revise{parameter values} are listed in \autoref{tab:table4}. 
\revise{The models are seeded with} 203,418 particles\revise{;} 153,752 dark matter particles, 36,756 disc particles and 12,910 bulge particles. 
\revise{When checking the convergence of the numerical resolution}, we \revise{multiplied this total particle number by five (to 1,017,090). }
\revise{Because the GalactICS code uses a Poisson solver,} the total mass \revise{in} our models is \revise{altered by changing the disc thickness. }
\autoref{tab:table3} \revise{lists the ratios of the rotation velocity to the velocity dispersion of the stellar component and} the \revise{masses} of the bulge, disc and dark matter halo in each disc model. 
\revise{The effective radius of the constructed bulge ($\sim200$~pc) is consistent with the relationship between the radius and stellar mass \citep{Gadotti2009}. }

\begin{table}
 \centering
 \begin{minipage}{84mm}
  \caption{Mass \revise{abundances} of the progenitor models\label{tab:table3}}
  \begin{tabular}{@{}rccc@{}}
  \hline
  Model                & THIN &THICK &HOT\\\hline
  $Z_{\rm{d}}$ [kpc]   & 0.13 & 0.52 & 1.11\\
  $v_{\rm{max}}/\sigma$& 10.8 & 2.5 & 1.3\\
  Bulge [$M_{\sun}$]   & 2.9$\times$ $10^8$ & 3.1$\times$ $10^8$ & 3.1$\times$ $10^8$\\
  Disc [$M_{\sun}$]    & 7.8$\times$ $10^8$ & 7.3$\times$ $10^8$ & 6.5$\times$ $10^8$\\
  DM halo [$M_{\sun}$] & 3.2$\times$ $10^9$ & 3.5$\times$ $10^9$ & 3.9$\times$ $10^9$\\
\hline
\end{tabular}
\end{minipage}
\end{table}

\begin{figure}
  \includegraphics[width=\columnwidth]{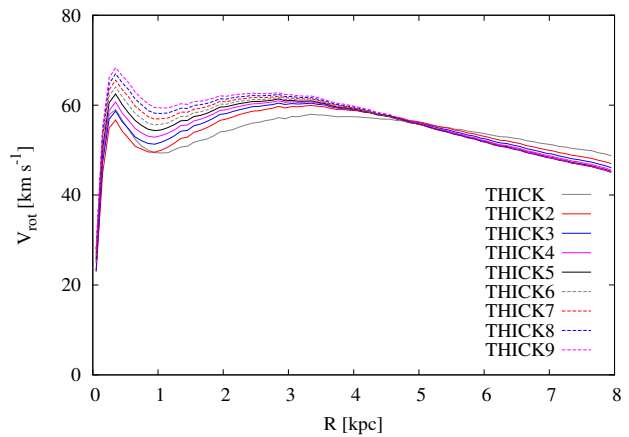}
  \caption{\revise{Rotation curves in various disc models (THICK (lowest disc rotation velocity) to THICK9 (highest disc rotation velocity)). }
 \label{figure_rotfull}
  }
\end{figure}

\revise{To expand the parameter space of the THICK model, we fix $Z_{\rm d}$ and vary the rotation curve, which would affect the shape of the GSS. }
\revise{The} different rotation velocity of the disc \revise{is simply varied by changing the} central gravitational potential of the dark matter halo\revise{. The resulting models are named THICK2--THICK9 (see \autoref{tab:table4}). }
\revise{Fig.~\ref{figure_rotfull} shows rotation curves of the various disc models. }
\revise{The THICK disc models are numbered in order of increasing rotation velocity of the disc inner 5~kpc. }
\revise{All of the progenitor models are consistent with the observed baryonic Tully--Fisher relation in the scatter range \citep{McGaugh2000}. }

\revise{In the $N$--body simulations, the gravity is calculated by an} original parallel tree code with a tolerance parameter of $0.5$\revise{. The Plummer} softening length \revise{is} $\sim 8$~pc\revise{. These parameters sufficiently resolve the bulge component and reduce numerical two-body relaxation. }
The orbit \revise{is integrated by} a second-order leapfrog integrator \revise{with a shared timestep of approximately 10 kiloyears}. 
\revise{
When testing the convergence of the numerical resolution, we perform an additional high-resolution run using the same code. 
As mentioned above, the total particle number in the high-resolution run (1,017,090) is five times that of the normal resolution. 
Additionally, we increase the particle number to 16,777,216 ($\sim16$ times that of the high-resolution) and conduct a highest-resolution run. 
Besides verifying convergence, the highest-resolution run reveals the metallicity distribution in the faint regions such as the broad western structure of the GSS, by which we reduce the Poisson noise in the $N$--body simulations. 
In the highest-resolution model, we employ the gravitational octree code optimized for Graphics Processing Units (GPU), GOTHIC (\citeauthor{Miki2016b} in prep.), which implements a hierarchical time step with a second-order Runge--Kutta integrator and the multipole acceptance criterion proposed by \citet{WarrenSalmon1993} and \citet{SalmonWarren1994}. 
The accuracy control parameter $\Delta_{\rm acc}$ is set to be $2^{-8}$. 
The resolutions and codes of the normal, high and highest resolution models are summarised in \autoref{tab:table_resolution}. 
}
Numerical calculations \revise{are} carried out on \revise{the} T2K--Tsukuba System, HA--PACS System, COMA System and a workstation \revise{at the} Center for Computational Sciences, University of Tsukuba\revise{, Japan}. 

\begin{table}
 \centering
 \begin{minipage}{84mm}
  \caption{Resolution \revise{(particle number $N$)} and \revise{programming} codes \revise{at different resolutions}\label{tab:table_resolution}}
  \begin{tabular}{@{}cccc@{}}
  \hline
  Resolution & Normal & High & Highest\\
\hline
$N$ & 203,418 & 1,017,090 & 16,777,216\\
purpose& Large parameter &Convergence test &High-quality\\
       & survey          &for large survey &data analysis\\
code   & Original tree   &Original tree    &GOTHIC \\
       & on CPUs         &on CPUs          &on GPU \\
section& \S \ref{sec:NR} & \S \ref{sec:conv} & \S \ref{sec:metal}--\S \ref{sec:extended_shell}\\
\hline
\end{tabular}
\end{minipage}
\end{table}

\subsection{Rotation of satellite progenitor}

\begin{figure}
  \includegraphics[width=\columnwidth]{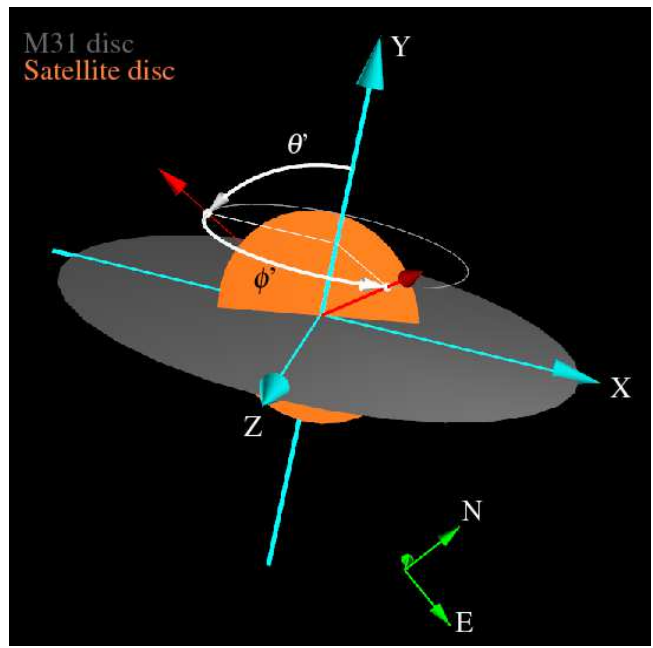}
  \caption{Initial inclination $(\phi',\theta')$ of the spin axis of the progenitor's disc \revise{in the specified coordinate system}. 
Gray elliptical disc \revise{delineates} the M31's disc \revise{in the numerical simulation coordinates} and the orange elliptical disc is the inclined disc of the progenitor. 
The green compass shows the north, east and the \revise{Local Standard of Rest directions}. 
The Earth \revise{locates at} the backside of M31's disc. 
Three-dimensional visualisation was conducted with the S2PLOT programming library \citep{Barnes2006}. 
    \label{fig_ini}
  }
\end{figure}

\revise{To reproduce the observed shapes of the GSS and shells, we require} an initial inclination of the disc progenitor\revise{, which is precluded in spherical progenitor models. }
Therefore, \revise{we systematically vary} the initial inclination of the disc \revise{and simulate} galaxy merger between M31 and the progenitor galaxy. 
\revise{In this survey, we construct a two-dimensional plane in spherical coordinates} $(\phi',\theta')$ (later \revise{altered to} $\phi (\equiv 180\degr-\phi')$, $\theta (\equiv 180\degr-\theta')$) \revise{(as shown in Fig.~\ref{fig_ini})}. 
Initially, the pole of the satellite system \revise{aligns with} the direction of the angular momentum vector of \revise{M31's disc}. 
In other words, \revise{when of $(\phi',\theta')=(0\degr,0\degr)$,} the disc \revise{planes} and the spin \revise{axes} of the progenitor \revise{and M31 discs are identical. }
\revise{The X(Z) axis corresponds to the minor (major) axis of \revise{M31's disc} and the arrow points to the north-western (eastern) side \revise{of M31's centre}. }
The polar angle $\theta'$ \revise{ranges from} the pole ($+$Y) direction to \revise{the} $-$X direction \revise{($0\degr\lid \theta'<180\degr$)} and the azimuthal angle $\phi'$ is measured anti-clockwise from \revise{the} $-$X axis on the disc plane \revise{in the} $+$Z direction \revise{($0\degr\lid \phi'<360\degr$). }
\revise{The} spin axis of the disc dwarf galaxy is inclined \revise{first} by $\theta'$ and then by $\phi'$. 

\revise{Initially, the disc of the} progenitor galaxy \revise{is} inclined by $(\phi,\theta)$\revise{, and the} $(\phi,\theta)$ \revise{plane is comprehensively surveyed with a} grid width of $30\degr$. 
Next, we carefully \revise{examine parameter spaces that reasonably reproduce the GSS. }
In each disc model \revise{(THIN, THICK and HOT)}, \revise{we simulate} approximately 350 runs on the $(\phi,\theta)$ plane \revise{and identified the appropriate parameter spaces on that plane (e.g. Fig.~\ref{peak_east}). }
\revise{
In the beginning, we explore the case of the models THIN, THICK, and HOT and find an appropriate range of the parameter space on the plane $(\phi,\theta)$ (e.g., Fig.~\ref{peak_east}). 
We then perform detailed simulations in the appropriate parameter spaces ($-50\degr\leq\phi\leq60\degr$ and $10\degr\leq\theta\leq105\degr$). 
All models are iterated through 74 runs except the THICK7 model (176 runs). 
}

\section{Numerical Results}
\label{sec:NR}
\subsection{Representative models}
\label{sec:typical}

\begin{figure}
  \includegraphics[width=\columnwidth]{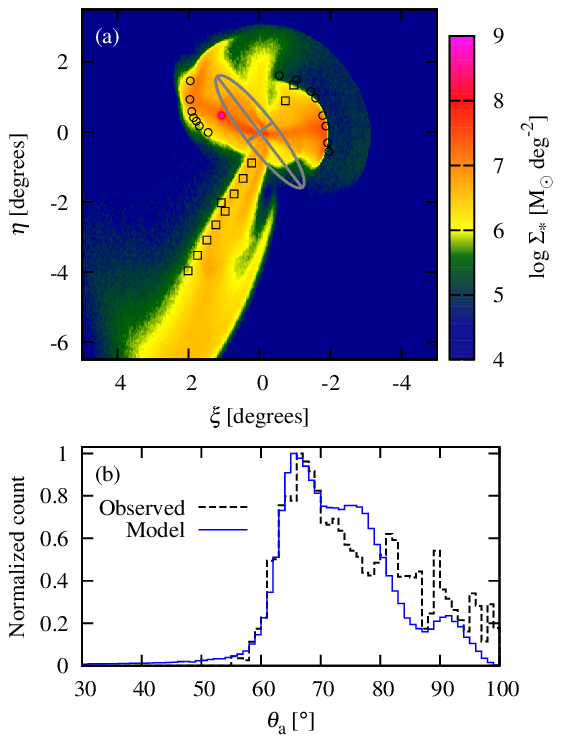}
  \caption{(a) Surface mass-density distribution of the disrupted progenitor\revise{, under the parameters yielding} one of the most successful results (model THICK7 with $(\phi,\theta)=(-15\degr, 30\degr)$). 
The inclined elliptical line describes the shape of M31's disc. 
Square symbols \revise{indicate} the observed fields of the GSS \revise{\citep{McConnachie2003,Guhathakurta2006}}. 
Circles are the edge positions of the eastern and western shells \revise{analysed} by \citet{Fardal2007}. 
(b) \revise{Normalised} stellar count in the GSS as a function of azimuthal angle. 
Blue solid \revise{and black lines present} the $N$--body simulation \revise{results} and the observed profile of the GSS\revise{, respectively}. 
 \label{best--fitting}
  }
\end{figure}

\revise{In this section, we simulate galaxy collisions} between M31 and a disc dwarf satellite galaxy. 
\revise{First we demonstrate} a successful model that well reproduces the GSS axis, width of the sharp eastern edge and width of the broad western structure of the GSS. 
Fig.~\ref{best--fitting} shows the whole spatial distribution of the satellite galaxy \revise{remnant} and the normalised stellar number count\revise{,} as function of the azimuthal angle for one of the successful parameters (model THICK7 with $(\phi,\theta)=(-15\degr, 30\degr)$). 
\revise{To ensure a high-quality analysis, we here use the data of the highest-resolution runs (see \autoref{tab:table_resolution}). }
The \revise{shapes and sizes} of the north-eastern \revise{and western shells} are also well reproduced. 
\revise{The westernmost side of the GSS exhibits an arm-like structure rather than the observed smooth structure. 
However, the true structure is poorly understood because the GSS features are faint in that region and substructures have been observed there. 
The northern side of the eastern shell presents a dense region at the approximate position of the northern spur reported by \citet{Ferguson2002}. 
The faintest structures are an extended shell structure outside the western shell (\S \ref{sec:extended_shell}) and a spillover at the tip of the GSS ($\eta \sim -5\degr$ and $\xi \sim 4\degr$). 
\citet{Ibata2007} reported a similar structure called Stream B, which is mainly composed of metal-rich stars. 
Whereas the observed structure is almost perpendicular to the GSS, the simulated one has a smaller interior angle. }

\begin{figure*}
  \includegraphics[width=.99\textwidth]{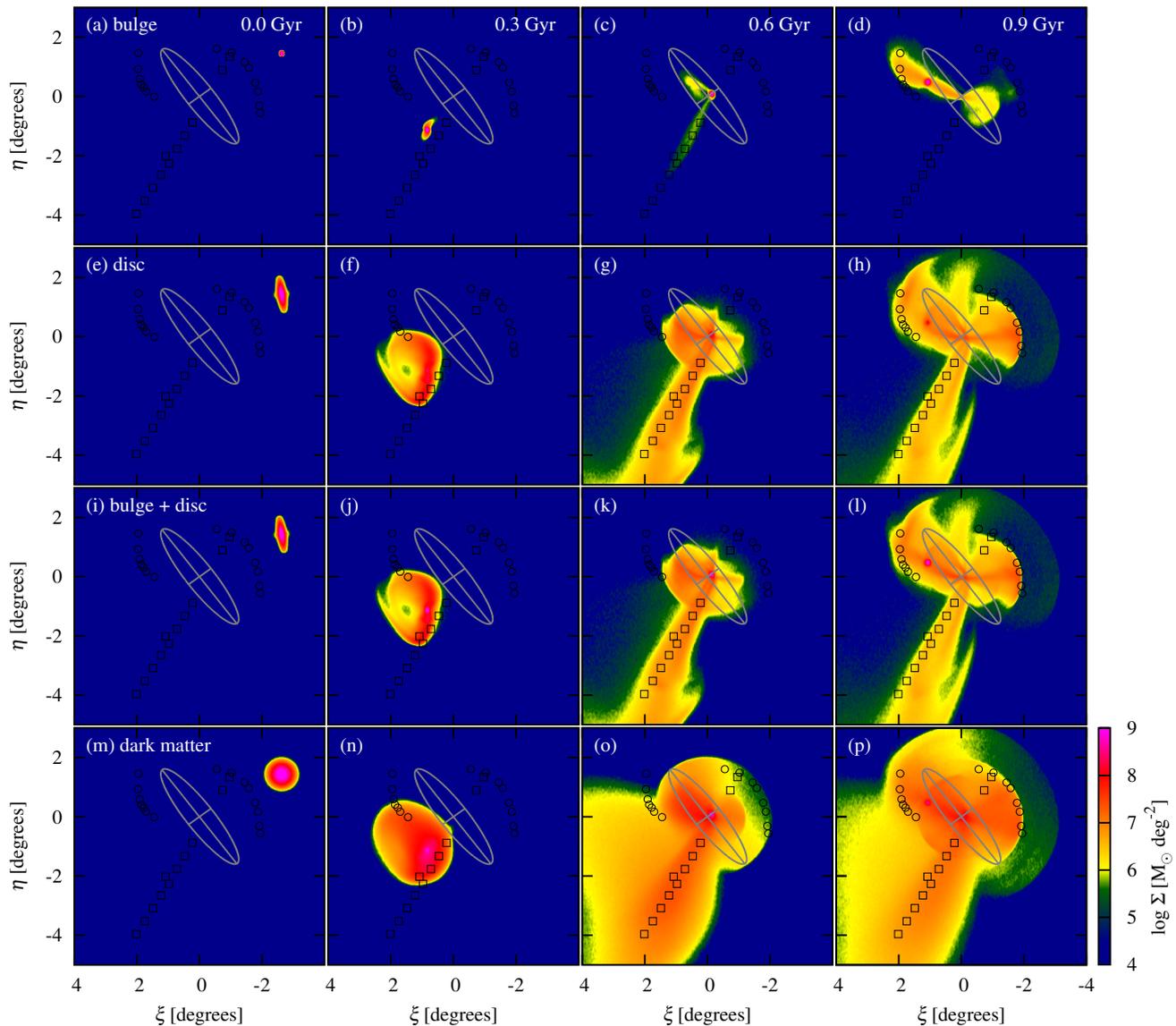}
  \caption{Evolution of the surface mass-density distribution of the disrupted dwarf galaxy \revise{in sky coordinates}. 
    \revise{Top to bottom panels show the evolution} of the bulge component (a--d), disc component (e--h), bulge and disc components (i--l) and dark matter halo component (m--p). 
    \revise{Left} to right panels \revise{present the mass-density distributions at} 0.0~Gyr, 0.3~Gyr, 0.6~Gyr and 0.9~Gyr (current epoch) after the start of the simulation run. 
    Symbols and \revise{lines} in each panel are \revise{those of} Fig.~\ref{best--fitting}a. 
    \label{evolution}}
\end{figure*}

Fig.~\ref{evolution} shows the time evolution of the disrupted dwarf galaxy on the sky coordinate \revise{up to} current epoch. 
The progenitor galaxy\revise{, initially located in} the north-western area of M31 (Fig.~\ref{evolution}i)\revise{, collided} almost head-on with M31. 
\revise{From its} initial condition\revise{, the progenitor disc reached} the first pericentric passage of the orbital motion \revise{in approximately one dynamical time of the progenitor's disc}. 
\revise{The progenitor is then} disrupted and \revise{a} component spreads into \revise{the} south-eastern area \revise{with growth} the GSS (Fig.~\ref{evolution}j). 
\revise{Simultaneously,} part of the debris falls and \revise{enters the} western side of \revise{M31's centre. After} the second pericentric passage\revise{, the debris spreads into a wide fan} called the eastern shell (Fig.~\ref{evolution}k). 
Immediately, some of \revise{the debris moves} to the western area\revise{, forming a} similar shell called the western shell (Fig.~\ref{evolution}l). 
In each run, \revise{the} current epoch is defined as the snapshot\revise{, in which} the simulated edge positions of the eastern and western shells \revise{best} match the observed positions. 

In Fig.~\ref{evolution}l, the dense region of the simulated GSS lies along the observed fields \revise{(indicated by open squares)} \citep{McConnachie2003,Guhathakurta2006}. 
\revise{In the simulation results, the} boundaries of the shell structures at the north-eastern and western \revise{areas} reach the observed edges of the shells \revise{analysed} by \citet{Fardal2007} \revise{(indicated by black circles)}. 
\revise{Most} of the bulge component resides in the eastern shell (Fig.~\ref{evolution}d)\revise{; almost none is found} in the GSS area. 
\revise{This occurs because the} bulge stars are \revise{strongly bound} by the gravitational potential of the progenitor's bulge \revise{and therefore survive} the tidal disruption of \revise{M31's} gravitational field at the first pericentric passage. 
\revise{The simulated} progenitor also \revise{exhibits} a spherical dark matter halo component, \revise{which evolves as shown in} the bottom panels of Fig.~\ref{evolution}. 
The spatial distribution spread over \revise{a} quite broad region around M31\revise{, especially at} the eastern side of the GSS. 
Interestingly, the sharp edge of the dark matter distribution in the western M31 \revise{appears to locate at} the stellar component. 

\begin{figure}
  \includegraphics[width=\columnwidth]{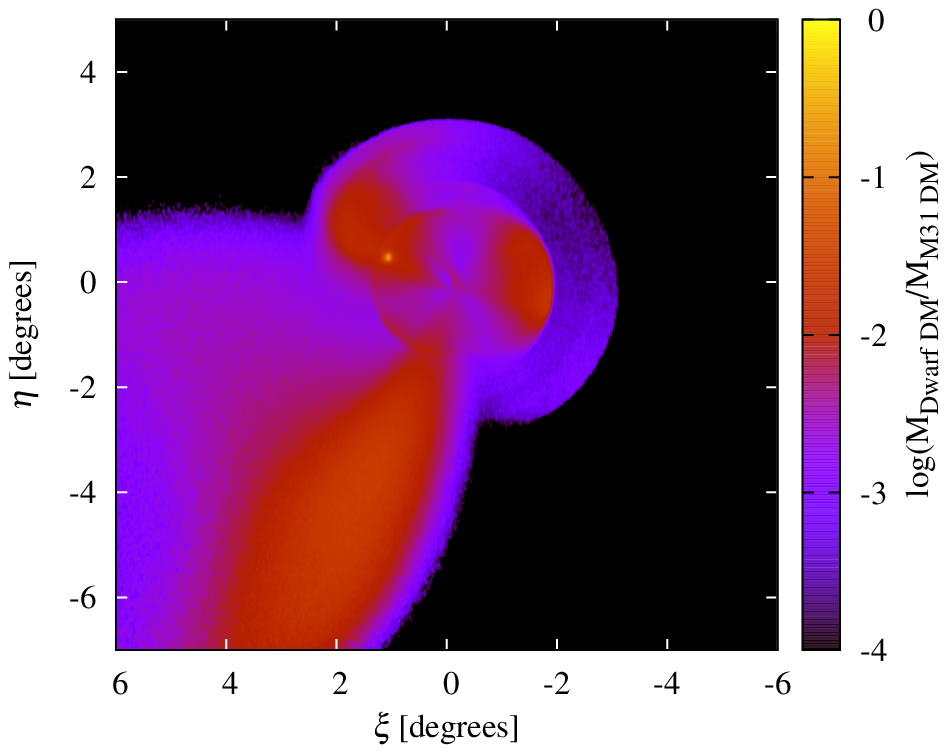}
  \caption{Mass ratio \revise{of} the dark matter halo initially \revise{inhabited} with the progenitor galaxy\revise{, relative to} the dark matter halo of M31. 
\revise{The inclination of the disc model is $(\phi,\theta)=(-15\degr, 30\degr)$ (model THICK7). }
    \label{fraction_dm}
  }
\end{figure}

\revise{
Fig.~\ref{fraction_dm} shows the mass fraction of the dark matter component originally surrounding the progenitor dwarf, relative to M31. 
The assumed mass distribution of the dark matter halo inhabited with M31 is described in \S \ref{sec:M31model}. 
This figure is constructed from the snapshot of Fig.~\ref{evolution}p. 
The highest mass fraction is approximately 0.1. 
Relative to the smooth dark matter halo of M31, the progenitor's dark matter is significantly enhanced at the edge of the western shell. 
Interestingly, this region of mass-density enhancement coincides with the stellar structure at the western shell. 
However, this result indicates that the mass fraction is too small to detect even with TMT (thirty-metre-telescope), which exploits the weak lensing of the background halo stars. 
}

\begin{figure*}
  \includegraphics[width=.99\textwidth]{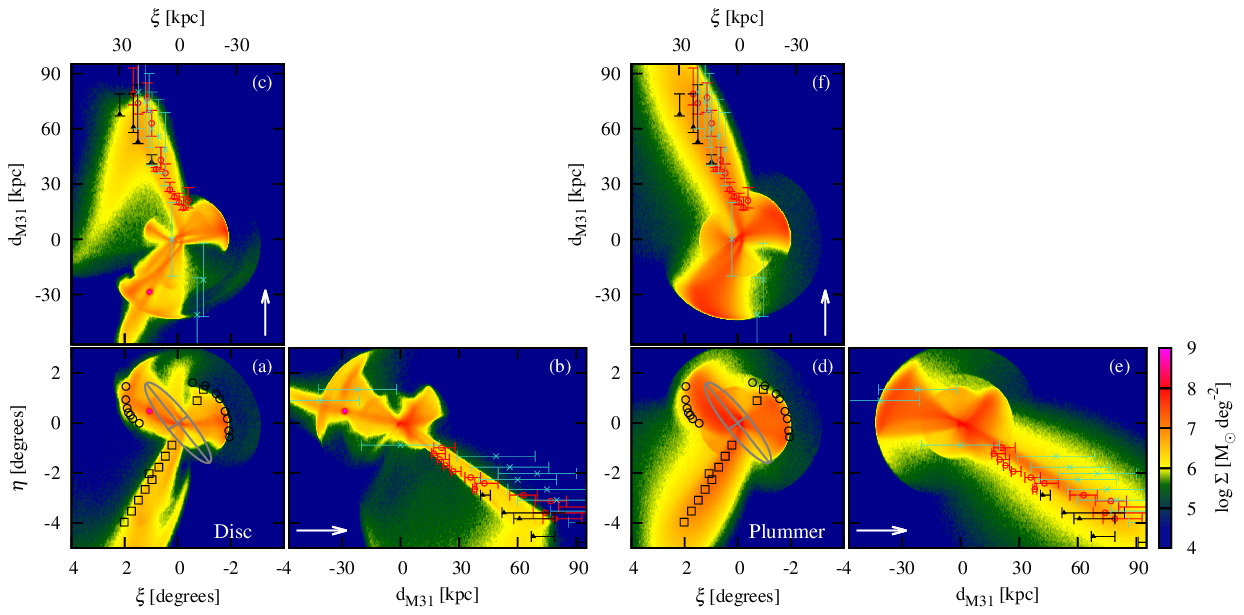}
  \caption{3D distribution of the colliding satellite galaxy \revise{in} the disc \revise{model} (both disc and bulge stars are plotted) and \revise{the} spherical progenitor \revise{model} at the best-fitting epoch. 
    Symbols and \revise{lines} in each \revise{graphic} are \revise{those of} Fig.~\ref{best--fitting}a. 
    Observed distances \revise{in} each region are \revise{indicated as follows:} open red circles with error bars (best parameters in \citet{Conn2016}), filled black triangles (most likely parameter values in \citet{Conn2016}) and cyan crosses \citep{McConnachie2003}.
    Left panels (a--c): \revise{Results of the successful} disc model \revise{(progenitor model using the same parameters as Fig.~\ref{evolution}l). }
    Right panels (d--f): Results \revise{of} the spherical symmetric Plummer model. 
    Panels (a) and (d): View on the sky \revise{coordinates}. 
    Panels (b) and (e): View on the line-of-sight depth $\rm{d_{M31}}$ [kpc] \revise{versus} $\eta$ [degrees] \revise{plane}.
    Panels (c) and (f): View on the line-of-sight depth $\rm{d_{M31}}$ [kpc] and $\xi$ [degrees] \revise{plane}.
    White arrows show the line-of-sight direction from Earth. 
    \label{disk_plummer3d}}
\end{figure*}

Fig.~\ref{disk_plummer3d} displays the 3D distribution of the debris in the \revise{spherically} symmetric and axisymmetric progenitor models. 
\revise{The merger with a spherical galaxy is simulated under the same mass-resolution (11,520,000 particles) as the baryonic component of the disc model in the highest-resolution run. }
The spherical dwarf satellite galaxy has a Plummer equilibrium distribution with \revise{a} total mass of $2.2\times 10^9M_{\sun}$ and \revise{a} scale radius of $1.03$~kpc. 
\revise{The three}-dimensional distribution of the simulated GSS is consistent with the \revise{latest observations} by \cite{Conn2016}. 
The simulated GSS is \revise{slightly} shorter than the observed GSS, \revise{possibly because we imposed} an artificial radial cutoff in the initial progenitor's disc for simplicity. 
\revise{Dynamically} cold components (e.g. fine structures \revise{in} the inner \revise{regions} of \revise{the} eastern and western shells) \revise{appear in the disc merger scenario (Fig.~\ref{disk_plummer3d})}. 
The most important difference is the GSS \revise{width; in particular,} a dynamically cold component on the western side of the GSS. 
\revise{The southern and western spread of the GSS is much broader in the disc merger than in the spherical progenitor merger in the 3D view. }

\begin{figure*}
  \includegraphics[width=.99\textwidth]{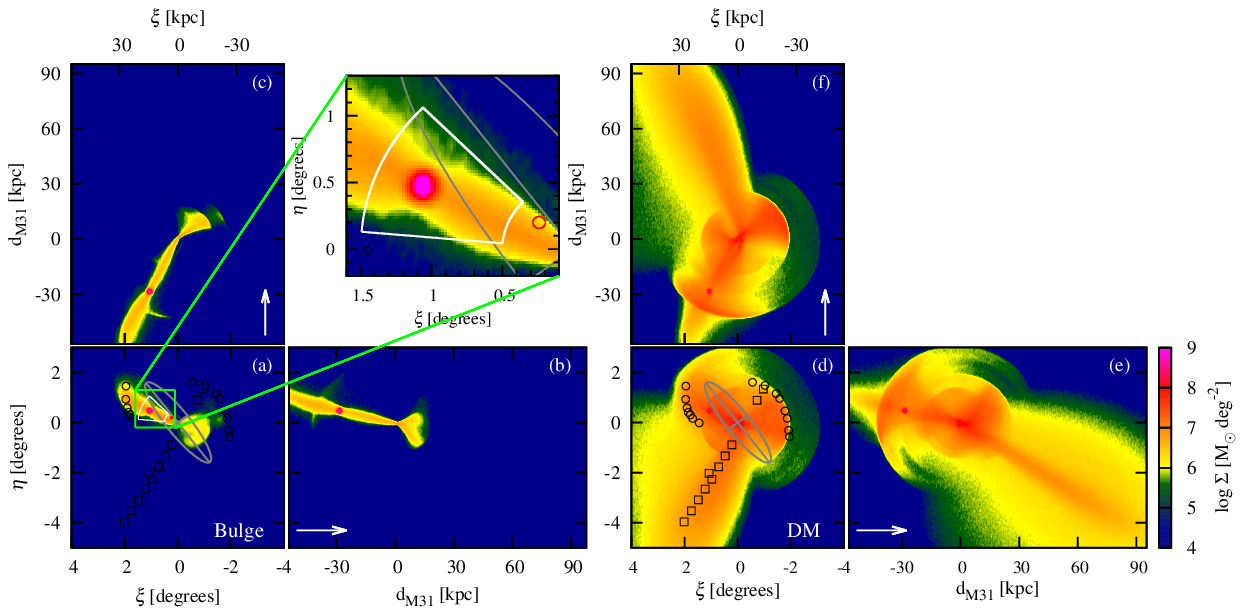}
  \caption{3D distribution of the bulge (left panels) and dark matter (right panels) components of the disrupted dwarf galaxy at the best-fitting epoch. 
    Symbols and \revise{lines} in each panel are \revise{those of} Fig.~\ref{best--fitting}a. 
    Viewing angle is the same as in Fig.~\ref{disk_plummer3d}. 
    The upper right \revise{graphic in} panel (a) \revise{enlarges the} 1.5\degr$\times$1.5\degr~region \revise{outlined in green}. 
    \revise{The} phase-space distribution of the bulge component \revise{is analysed in the white-outlined region}(see \S \ref{sec:MBH_bulge}). 
    The red circle \revise{indicates} the position and size of a recently discovered density enhancement on M31's disc \citep{Davidge2012}. 
    \label{disk3d}}
\end{figure*}

Fig.~\ref{disk3d} shows the 3D \revise{distributions} of the bulge and the dark matter halo components at the best-fitting epoch. 
In \revise{panels b and c,} the progenitor's bulge is elongated \revise{along} the line-of-sight direction \revise{by} the tidal force of M31's potential. 
\revise{Several shell-like structures appear in the} 3D view of the dark matter halo component of the disrupted progenitor\revise{, indicating that part of the} dark matter halo component crosses the central region of M31 several times. 

\begin{figure}
  \includegraphics[width=\columnwidth]{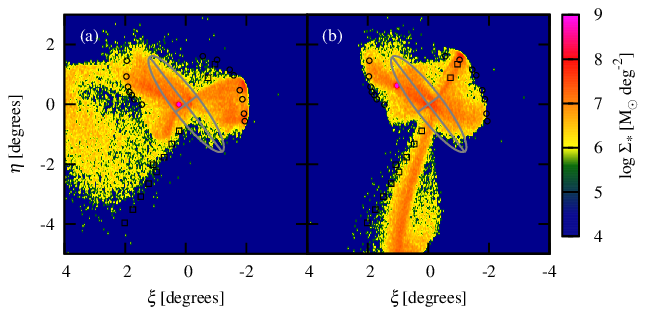}
  \caption{Projected stellar mass-density \revise{distributions of (a) an} initially clockwise-rotating disc model \revise{(model THICK with $(\phi,\theta)=(-90\degr,100\degr)$)} in the sky at the best-fitting epoch \revise{and} (b) \revise{an} initially anti-clockwise-rotating disc model \revise{(model THICK with $(\phi,\theta)=(90\degr,75\degr)$)}. 
    Symbols and \revise{lines} in each panel are \revise{those of} Fig.~\ref{best--fitting}a. 
    \label{figure5}
  }
\end{figure}

\revise{We now present some} typical \revise{results of clockwise- and anti-clockwise-rotating} disc models \revise{in} the sky coordinate \revise{system}. 
Figs~\ref{figure5}a and b show the stellar mass-density \revise{distributions} of almost clockwise ($(\phi,\theta)=(-90\degr,100\degr)$) and anti-clockwise ($(\phi,\theta)=(90\degr,75\degr)$) rotating disc models, respectively. 
\revise{The} shape of the debris \revise{largely differs} from \revise{that of} the galaxy\revise{--spherical dwarf merger}. 
The progenitor\revise{--M31 collision was} almost head-on, and the \revise{progenitor centre} passed \revise{barely east of M31's centre}. 
The shortest distance \revise{between the progenitor and M31 centres was $1$~kpc} at the first pericentric passage. 
In addition, the progenitor's disc is \revise{more} than $1$~kpc \revise{across} (\S \ref{sec:satellite_models}). 
\revise{As the} details depend on the inclination of the progenitor \revise{disc}, the main component of the progenitor passes \revise{to the east of} of \revise{M31's centre}, but a \revise{portion enters} the western area of \revise{M31's centre}. 

\revise{Clearly}, there is no GSS-like component in Fig.~\ref{figure5}a. 
Instead, \revise{we observe} a curious structure at the south-eastern area of M31. 
\revise{This occurs because} the stellar component \revise{passing to the west of M31's centre} at the first pericentric passage \revise{has a clockwise-rotating} component\revise{, so is ejected eastward}. 
On the other hand, \revise{in Fig.~\ref{figure5}b, the stellar component passing just east of M31's centre has an anti-clockwise-rotating component, so the GSS is slight on the eastern side but spreads broadly on the western side. }
However, the direction of the \revise{simulated} GSS \revise{offsets  relative to that of the observed GSS}. 

To \revise{summarise the above findings}, the direction \revise{and shape} of the GSS \revise{strongly depend on} the rotation of the progenitor galaxy. 
\revise{To} evaluate \revise{how} the initial inclination of the progenitor \revise{affects the GSS}, we \revise{analyse} the reproducibility of the GSS axis, the width of the eastern edge, and the width of the broad western extent \revise{in the following subsections}.

\subsection{GSS axis}

The \revise{direction of the} GSS axis \revise{explicitly informs} the reproducibility of the GSS. 
As stated in \S \ref{sec:typical}, the azimuthal angle of the density peak in the GSS \revise{largely} depends on the initial inclination of the progenitor's disc. 
\revise{To} examine the effect of \revise{this initial parameter}, we \revise{require} a complete sweep of the large parameter space. 

\begin{figure*}
  \includegraphics[width=.99\textwidth]{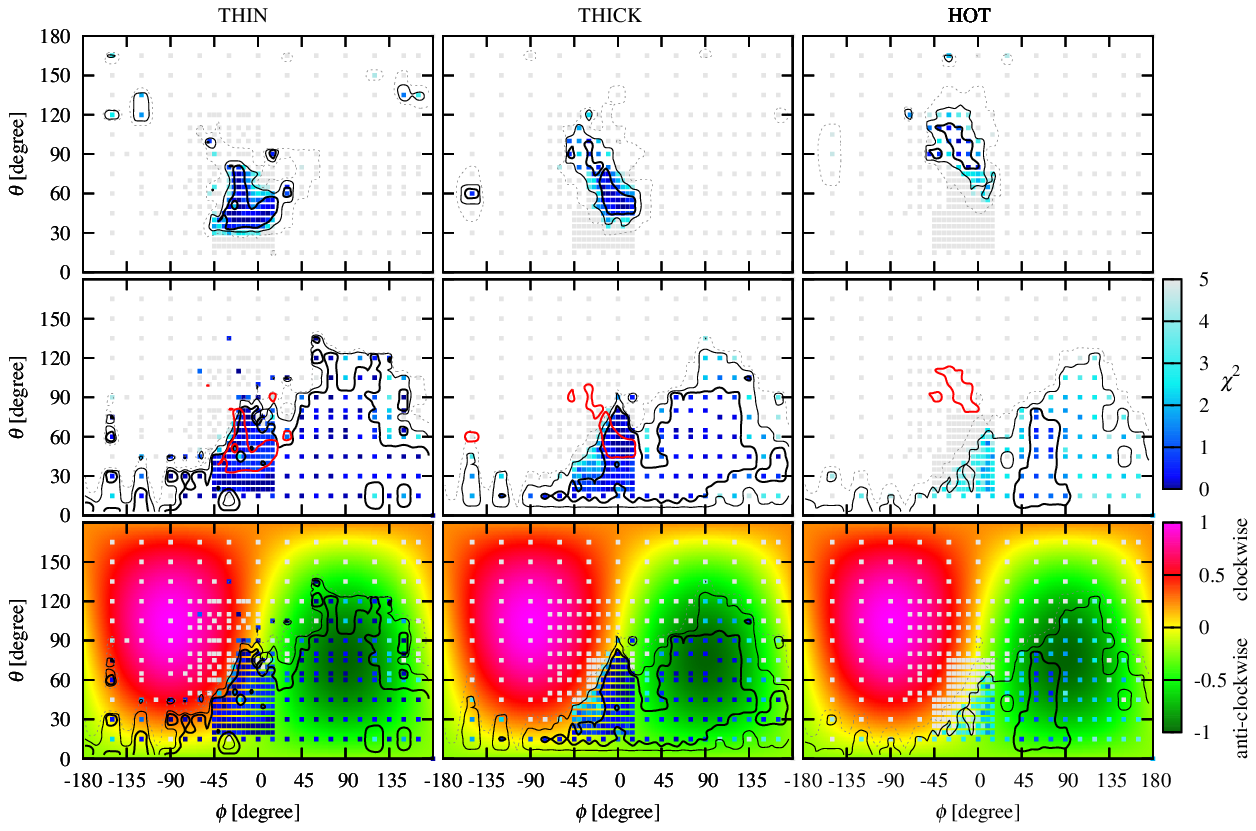}
  \caption{Top panels: \revise{Comparisons} between the azimuthal \revise{angles} of the density \revise{peaks} in the observed GSS and the simulation runs in each disc model on the plane $(\phi,\theta)$ of the initial inclination of the progenitor's disc \revise{($\chi^2$ analysis)}. 
    Middle panels: \revise{Comparisons} between the observed \revise{and simulated eastern edge widths}. 
    \revise{Coloured} squares show the $\chi^2$ \revise{values of} the simulated parameters. 
    Thick solid, thin solid and dashed contour \revise{lines indicate the} $1\sigma$, $2\sigma$ and $3\sigma$ confidence intervals of the $\Delta \chi^2$, respectively. 
    Thick red \revise{contours describe the} $1\sigma$ confidence \revise{intervals} of the azimuthal \revise{angles} of the density \revise{peaks} in the GSS \revise{(top panels)}. 
    Bottom panels: The initial spin axis of the progenitor's disc on the observed frame \revise{is related to} the eastern edge width of the GSS. 
    The \revise{colour} bar shows the inner product of the normalised line-of-sight vector into \revise{M31's centre} and the normalised initial spin vector of the progenitor's disc. 
    Magenta (green) \revise{areas indicate} clockwise (anti-clockwise) \revise{rotation of the progenitor's disc}. 
    \revise{Within} the yellow area, the disc is \revise{viewed} almost edge-on from the Earth. 
    \label{peak_east}}
\end{figure*}

\revise{To quantitatively compare} the observed \revise{and simulated GSS axes} $\theta_{\rm axis}$, we \revise{conduct a} $\chi^2$ analysis. 
\revise{Mimicking} the observed data \revise{analysis}, \revise{we obtain the simulated} GSS axis by an asymmetric exponential fitting of the azimuthal distribution of the GSS \revise{(as described in \S \ref{sec:asymmetry})}. 
\revise{The top} panels in Fig.~\ref{peak_east} show the $\chi^2$ \revise{maps of} the simulated GSS axis at the best-fitting epoch on the $(\phi,\theta)$ plane \revise{in each model}. 
\revise{Here we assume an observed uncertainty $\sigma_{\rm obs}^{\rm peak}$ of $\delta_{\rm obs}^-$, the observed width of the eastern edge (see \autoref{tab:table2})}. 
The bluer region well matches the observed GSS direction and the $(\phi,\theta)$ parameter space \revise{that reproduces} the observed GSS axis \revise{is tightly constrained}. 
The thick black line in each panel describes the 1$\sigma$ confidence interval of $\Delta \chi^2$ (=1)\revise{. The minimum $\chi^2$ value in the THIN, THICK and HOT disc models are 0.01, 0.28 and 0.44, respectively. }

The well-fitting area \revise{apparently shifts to} larger $\theta$ \revise{with increasing} scale height of the disc. 
\revise{The} reason for this \revise{trend is discussed} in \S \ref{sec:peak_shift}. 
Here \revise{the} $\chi^2$ value \revise{in the Plummer model is} 2.5, \revise{reasonably suitable for detecting} the GSS axis.

\subsection{Eastern edge of the GSS}
\label{sec:edge_reproduce}

The star counts sharply \revise{decrease} from the GSS axis \revise{in} the eastward direction\revise{, relative to the} western direction. 
We \revise{analyse} the eastern edge \revise{similarly} to the GSS axis. 

\revise{To} compare the \revise{observed and simulated widths} of the eastern edge, we \revise{conduct a} $\chi^2$ analysis assuming an \revise{observational uncertainty} $\sigma_{\rm obs}^{\rm east}\equiv\delta_{\rm{obs}}^-$ \revise{(the observed width of the eastern edge stated in \autoref{tab:table2})}. 
\revise{Mimicking} the observed data \revise{analysis}, \revise{we obtain} the width of the simulated GSS \revise{by an} asymmetric exponential fitting. 
The middle panels in Fig.~\ref{peak_east} show the \revise{simulated widths} of the \revise{GSS} eastern edge at the best-fitting epoch on the $(\phi,\theta)$ plane in each disc model. 
\revise{The minimum $\chi^2$ values in the THIN, THICK and HOT disc models are 0.01, 0.04 and 0.18, respectively. }
\revise{Note that in the} THIN and THICK \revise{models}, a specific parameter range \revise{replicates} both the observed GSS axis and \revise{the} eastern edge of the GSS. 

\revise{At present, how the} surface brightness profile of the GSS \revise{becomes asymmetric is unclear}. 
Here, we examine the mechanism \revise{that forms} the eastern edge of the GSS on the plane $(\phi,\theta)$. 
The bluer region well fits the observed \revise{width of the GSS} eastern edge and a clear boundary of $\chi^2$ values separates the well-fitting parameters from the \revise{remaining} $(\phi,\theta)$ parameter space. 
\revise{The colour maps in} the bottom panels \revise{of Fig.~\ref{peak_east} show} the \revise{directions} of the initial inclination of the progenitor's disc\revise{, viewed} from the Earth. 
\revise{These maps are overlaid on the $\chi^2$ maps of the eastern edge width of the GSS. }
The \revise{colour} bar indicates \revise{the} inner product of the normalised line-of-sight vector into \revise{M31's centre} and the unit vector of the progenitor's disc spin. 
This figure describes the \revise{behaviour} of the rotating disc in the sky\revise{, providing intuitive information on the successful conditions}. 
The boundaries in the middle panels of Fig.~\ref{peak_east} \revise{resemble} the curve of the edge-on region \revise{in} the sky \revise{coordinates}. 
\revise{Viewed} from the Earth, the spin vector of the progenitor's disc switches \revise{between} positive \revise{and} negative \revise{as it crosses} the curve. 
In other words, the curve is the switching line of the clockwise and anti-clockwise rotation of the progenitor's disc in the sky. 

\revise{Although the clockwise-rotating disc models nicely replicate the GSS axis, they fail to reproduce the sharp eastern edge. 
The successful parameter space resides in the plane ($(\phi,\theta)$ with $-45\degr<\phi<30\degr$ and $30\degr<\theta<80\degr$ (THIN model) or $-20\degr<\phi<20\degr$ and $40\degr<\theta<70\degr$ (THICK model)). }
\revise{The} Plummer model \revise{fails to construct the edge} and the minimum $\chi^2$ value \revise{is} 9.9. 
\revise{This confirms that} spherical progenitor models \revise{cannot} reproduce the \revise{sharp} the eastern edge.

\subsection{Broad western structure of the GSS}

In contrast to the eastern side, the star counts \revise{at the western side} of the GSS \revise{gradually} decrease from the GSS axis to the westward direction. 
We \revise{investigate the} internal structure \revise{of this region} and the reproducibility of the broad western structure\revise{,} comparing \revise{the} simulated \revise{results} with \revise{the} observed data. 
For comparison, we set the half width of the broad western structure $\delta_{\rm{obs}}^+/2$ \revise{(see \autoref{tab:table2})} as the \revise{uncertainty} $\sigma_{\rm obs}^{\rm{west}}$ \revise{in} the $\chi^2$ analysis. 
\revise{This} large \revise{uncertainty} $\sigma_{\rm obs}^{\rm{west}}$ \revise{reflects the faint and noisy nature of the wider western region of the GSS. Consequently,} the observed distribution \revise{fitting might} overestimate the \revise{true} width of the broad western structure. 
Nevertheless, the western side of the GSS is \revise{significantly} wider than the eastern edge \revise{in reality}. 

\begin{figure*}
  \includegraphics[width=.99\textwidth]{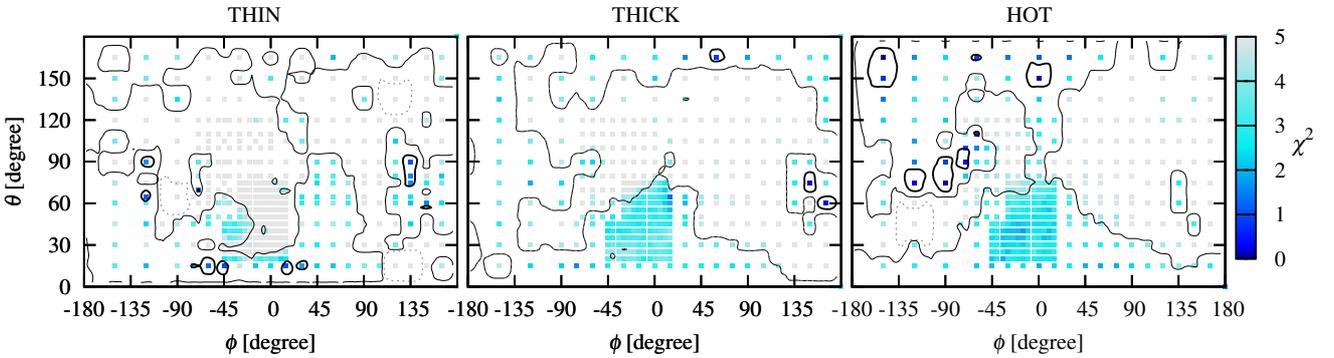}
  \caption{Same as Fig.~\ref{peak_east}, but \revise{comparing the observed and simulated} western \revise{widths} of the GSS. 
    \label{west}
  }
\end{figure*}

Fig.~\ref{west} shows the $\chi^2$ map \revise{of} the western side of the GSS at the best-fitting epoch \revise{in} the $(\phi,\theta)$ plane. 
The bluer region well reproduces the observed broad western width of the GSS. 
\revise{The minimum $\chi^2$ value in the THIN, THICK and HOT models are 0.74, 0.97 and 0.21, respectively. }
\revise{Again, we estimate} the width of the simulated GSS \revise{by an} asymmetric exponential fitting. 
\revise{This} analysis \revise{does} not \revise{apparently} limit the parameter space. 
\revise{In general, the thicker} the disc \revise{model}, the better the reproducibility of the observed western width. 
\revise{The best-fitting parameters in the} THICK and HOT \revise{models reproduce all three stellar distributions of the GSS:} the GSS axis\revise{,} the eastern edge \revise{and} the broad western structure. 

\begin{figure}
  \includegraphics[width=\columnwidth]{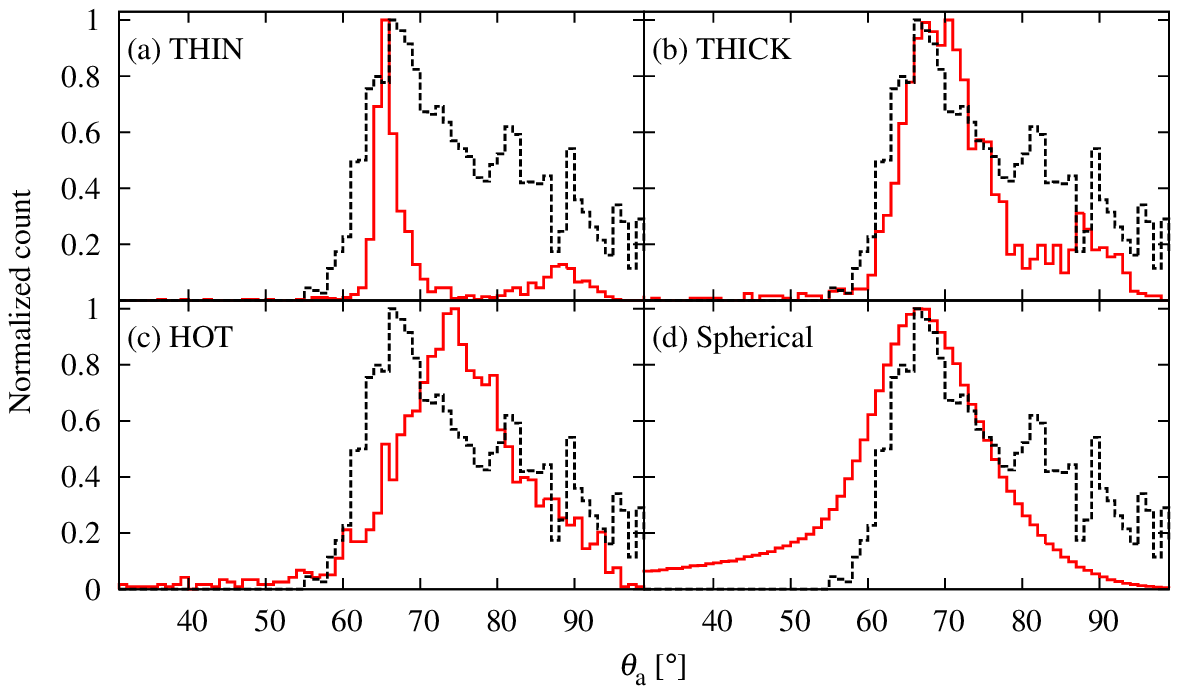}
  \caption{Azimuthal angle \revise{distributions (red lines) of the GSS simulated by the (a) THIN, (b) THICK, (c) HOT and (d) Plummer models. }
The inclination of the disc models is fixed \revise{at} $(\phi,\theta)=(5\degr,40\degr)$. 
\revise{The} observed \revise{GSS} distribution \revise{(black dashed lines) is also plotted in each panel}. 
    \label{histall}
  }
\end{figure}

Fig.~\ref{histall} \revise{plots} the normalised number count as a function of the azimuthal angle \revise{for $(\phi,\theta)=(5\degr,40\degr)$ in} various disc and spherical progenitor models. 
The selected \revise{parameters in Fig.~\ref{histall} yield successful results in the THICK model}. 
In the THIN \revise{model}, the azimuthal angle of the density peak in the GSS \revise{matches} the observed one\revise{, but} the stream \revise{is} dynamically too cold and \revise{is} too narrow. 
\revise{In contrast, the} THICK \revise{model} (Fig.~\ref{histall}b) well \revise{reproduces the} observed axis and eastern edge of the GSS. 
The western side of the simulated GSS \revise{is} wider than the eastern edge. 
\revise{At this inclination, the profile of the HOT model resembles that of} the spherical model (Fig.~\ref{histall}d)\revise{, with an angular shift} of the GSS axis. 
The western \revise{and eastern sides} of the simulated GSS \revise{are almost symmetric about} the GSS axis. 
\revise{In the Plummer model, the} $\chi^2$ value of the western width \revise{is} 3.2. 

\subsection{Effect of rotation velocity}
\label{sec:peak_shift}

As stated in \S \ref{sec:edge_reproduce}, the width of the eastern edge \revise{can be} explained by the anti-clockwise rotation of the progenitor's disc in the sky.
As displayed in Fig.~\ref{peak_east}, the \revise{parameters that well-fitted} the GSS axis \revise{are} substantially limited on the $(\phi,\theta)$ plane. 
\revise{When around $\phi=0$, the} parameter space that well reproduces the \revise{edge} width \revise{favours} smaller $\theta$. 
The THIN \revise{model reproduces} the GSS axis at smaller $\theta$ than the \revise{larger} scale height models. 
\revise{This} tendency \revise{might be attributable to the varying thicknesses} of \revise{the} disc models. 
\revise{In} fact, as \revise{mentioned} in \S \ref{sec:satellite_models}, \revise{changing the thickness of the disc model alters} the total mass of \revise{the} progenitor \revise{model, and hence the} rotation velocity \revise{of the} disc model. 
\revise{As varying the} rotation velocity of the progenitor's disc would shift the simulated GSS axis, we \revise{expand} the parameter space of the THICK \revise{model} (\revise{maintaining constant} $Z_{\rm{d}}$) \revise{to vary the} rotation curves. 

\begin{figure*}
\begin{center}
  \includegraphics[width=0.99\textwidth]{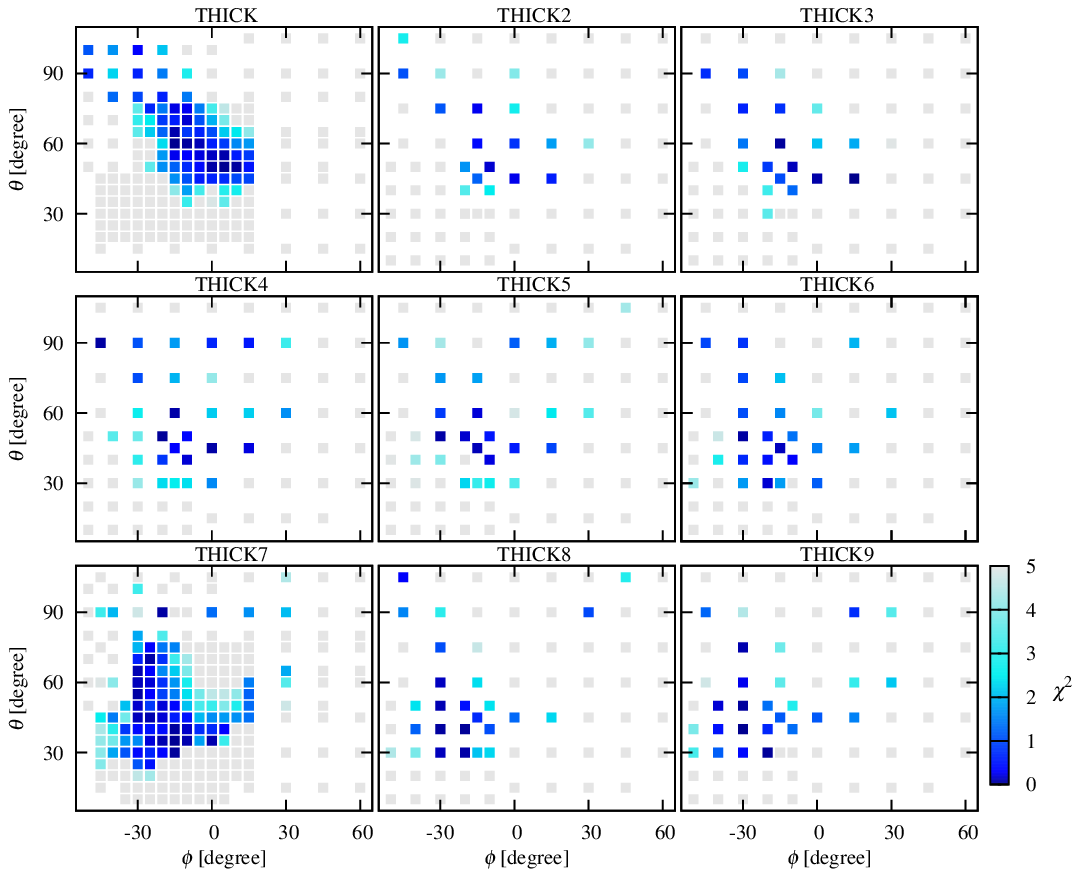}
  \caption{$\chi^2$ analysis \revise{of} the azimuthal angle of the \revise{GSS} density peak in the thick disc models. 
The symbols \revise{are explained in} Fig.~\ref{peak_east}. 
    \label{peak_shift}
  }
\end{center}
\end{figure*}

Fig.~\ref{peak_shift} \revise{presents the results} of the additional \revise{simulations in and around the acceptable parameter range of the THIN and THICK models. }
\revise{The number assigned to each disc model indicates its rotation speed (lowest for THICK, highest for THICK9; see \autoref{tab:table4}). }
\revise{Higher rotational velocity reduces the $\theta$ that reproduces the GSS axis. }
\revise{As the rotation speed of the disc increased, the fitted parameter space converges toward the THIN's parameter space (Fig.~\ref{peak_east} top-left panel). }

\revise{
Here, we briefly summarise our findings. 
A disc progenitor with higher rotation velocity shifts the GSS axis $\theta$ to smaller values. 
The eastern edge of the GSS forms only under anti-clockwise rotation of the progenitor dwarf, regardless of its disc thickness. 
In the extremely thin disc model, the debris is too dynamically cold to reproduce the observed spatial extent of the GSS. 
On the other hand, the HOT model could not form the observed sharp eastern edge. 
The broad structure at the western side of the GSS likely originates from an anti-clockwise rotating component passing just east of M31's centre at the first pericentric passage. 
}

\subsection{Line-of-sight velocity distribution}

\begin{figure}
  \includegraphics[width=\columnwidth]{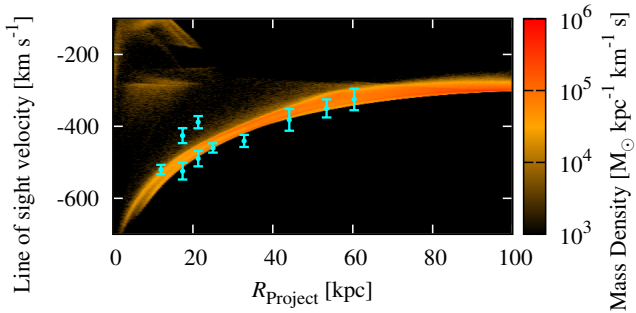}
  \caption{Mass density of the line-of-sight velocity distribution of the simulated GSS \revise{(model THICK7 with $(\phi,\theta)=(-15\degr, 30\degr)$)}. 
Cyan symbols \revise{are} observed data in each field \citep{Ferguson2004,Kalirai2006,Gilbert2009}. 
Each bar on the symbols \revise{indicates} the line-of-sight velocity distribution of the stars at \revise{that} distance. 
\label{velocity}
  }
\end{figure}

\revise{Additionally to} the photometric survey of stars\revise{, spectroscopic measurements of the line-of-sight velocity distribution in the GSS have also been carried out \citep{Ferguson2004,Kalirai2006,Koch2008,Gilbert2009}. }
The observed area \revise{almost follows} the extending direction of the GSS \citep{Ibata2004}. 
Fig.~\ref{velocity} shows the line-of-sight velocity distribution of the simulated GSS at the best-fitting epoch. 
The simulated data \revise{are those of the highest}-resolution run described in \S \ref{sec:typical} (model THICK7; $(\phi,\theta)=(-15\degr, 30\degr)$) \revise{and the simulation area is $30\degr<\theta_a<100\degr$ (see Fig.~\ref{figure2}a)}. 
\revise{The distance and} heliocentric systemic line-of-sight velocity of \revise{M31 are assumed as $780$~kpc and} $-300$ km s$^{-1}$\revise{, respectively} \citep{Font2006}. 
\revise{The simulated} line-of-sight velocity \revise{distributions are} consistent with that of the observed data. 

\revise{
\citet{Gilbert2009} reported an additional cold component (R$\lesssim$20~kpc) near the eastern edge of the GSS. 
This component is absent on the sky coordinates in our result (Fig.~\ref{disk_plummer3d}a), but a similar structure overlaps on the eastern shell (see the 3D map (($\xi, {\rm d_{M31}})=(0.5\degr, 5 {\rm kpc})$) in Fig.~\ref{disk_plummer3d}c). 
\citet{Miki2016} reported a similar component (third shell) in many parameter sets. 
}

\section{Discussions}
\label{sec:discussions}
\subsection{\revise{Availability} of the model and \revise{assumptions}}
\label{sec:conv}

\begin{figure}
  \includegraphics[width=\columnwidth]{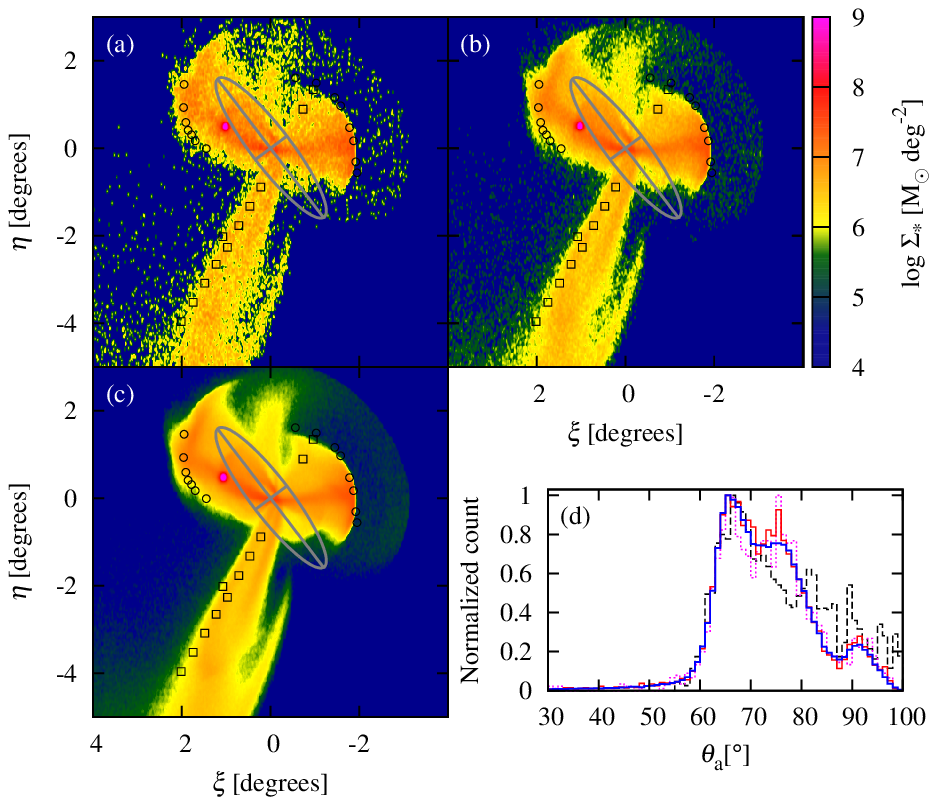}
  \caption{Convergence test of our \revise{simulations} with a successful parameter \revise{set} (model THICK7 with $(\phi,\theta)=(-15\degr, 30\degr)$). 
    \revise{Spatial distributions of the disrupted progenitor at the best-fitting epoch with different resolutions: 
    (a) normal-resolution simulation in parameter surveys, 
    (b) high-resolution simulation, which uses five times more particles than the normal-resolution runs, 
    (c) highest-resolution simulation, which is implemented in different code (GOTHIC). }
    Symbols and \revise{lines} in panels \revise{(a), (b) and (c)} are \revise{those of} Fig.~\ref{best--fitting}a. 
    \revise{(d)} Azimuthal angle distribution of the GSS. 
    The dotted magenta, \revise{solid red} and solid blue lines \revise{are generated at normal, high and highest resolutions}, respectively. 
    The black dashed line \revise{denotes} the observed data \citep{Irwin2005}. 
    \label{highreso}
  }
\end{figure}

\revise{To test the} convergence \revise{of the numerical resolution}, we \revise{perform a high}-resolution run \revise{using the parameters} that well \revise{reproduce} the observed structures (THICK7; $(\phi,\theta)=(-15\degr, 30\degr)$). 
The \revise{high}-resolution run \revise{simulates} 1,017,090 \revise{particles (five times the particle number in normal-resolution runs)}. 
Fig.~\ref{highreso} shows the convergence \revise{results}. 
\revise{Panel c presents the results of the highest-resolution run, which the number of particles is about 16 times higher than in the high-resolution run. }
\revise{All three runs yield similar global structures, but the distribution is somewhat noisy} in Fig.~\ref{highreso}a. 
\revise{The azimuthal angle distributions of the GSS are consistent in the normal- and higher-resolution runs (Fig.~\ref{highreso}d). }
Therefore, we consider \revise{the resolution of} our systematic surveys \revise{is sufficient}. 

\revise{\citet{Miki2014} systematically evaluated the infalling orbit of a spherical progenitor galaxy. They limited the possible orbital parameters within a narrow range including the orbit proposed by \citet{Fardal2007}.} 
\revise{The tight constraint originates from the strength of the tidal force exerted by M31's bulge and the passage duration of M31's central region. 
To form the GSS and both shells, the progenitor must be almost entirely disrupted. Therefore, assuming that the feasible orbital parameters are independent of the progenitor's morphology, we here adopt the orbit found by \citet{Fardal2007}. 
Our hypothesis will be tested in a future study (i.e. a systematic orbit survey of the disc progenitor). }

\citet{Kirihara2014} examined the outer density distribution of the dark matter halo of M31. 
\revise{They highlighted the necessity of reproducing the observed surface brightness ratios among the GSS and both shells. }
\revise{They suggested that the varying gravitational potential of M31 changes the evolutional timescale of the merger remnants and forms appropriate structures. }
However, \revise{in a spherical progenitor merger,} their best-fitting parameter \revise{did not replicate} the characteristic asymmetric structure of the GSS. 
In addition, \revise{their results might depend on} the morphology of the progenitor. 
For this reason, we \revise{first assume the generally adopted} conditions \revise{in our M31 model}. 
\revise{On the other hand, our successful disc model does not explain the observed surface density ratio. 
Therefore, the mass-density profile of M31's dark matter halo, the orbital initial conditions and the detailed morphology of the progenitor are interesting future investigations. }

\subsection{Metallicity distribution}
\label{sec:metal}

\begin{figure*}
  \includegraphics[width=.99\textwidth]{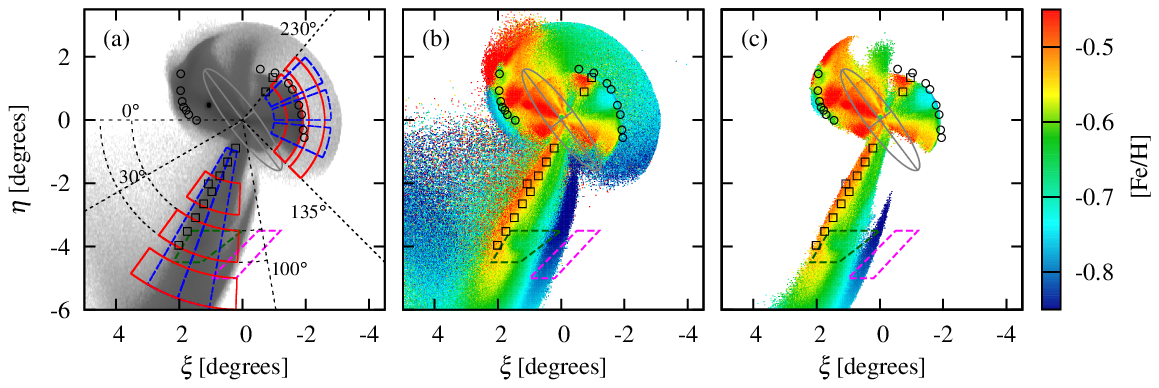}
  \caption{(a) Selected analysis \revise{regions}. 
    The background is \revise{the grey-scale} mass-density distribution of the disrupted progenitor galaxy \revise{(model THICK7 with $(\phi,\theta)=(-15\degr, 30\degr)$)}. 
    (b) Metallicity distribution \revise{with a} metallicity gradient of $\Delta$\feh$=-0.5$ and the mean metallicity of \mfeh$=-0.57$. 
    (c) Metallicity distribution filtered by the detection limit of INT/WFC. 
    Green and magenta dashed lines \revise{respectively denote the `core' and `cocoon' regions \citep{Ibata2007}}. 
    Other symbols and \revise{lines in each panel are indicated} in Fig.~\ref{best--fitting}a. 
    \label{metal_spatial}
  }
\end{figure*}

The \revise{GSS exhibits non-uniform} metallicity \revise{in the perpendicular direction to the GSS axis} \citep{Ibata2007,Gilbert2009}. 
The observed mean \revise{metallicities in} the `core' region (high-brightness region; outlined \revise{by the} green dashed line in Fig.~\ref{metal_spatial}) and \revise{the} `cocoon' region (western envelope; outlined \revise{by the} magenta dashed line in Fig.~\ref{metal_spatial}) are \mfeh$=-0.54$ and \mfeh$=-0.71$, respectively. 
On the other hand, nearby dwarf galaxies \revise{exhibit radial metallicity gradients} of $-0.6<\Delta$\feh$<0.2$ \citep{Koleva2009}\revise{. The radial metallicity is computed by: }
\begin{equation}
\Delta \feh \equiv\frac{{\rm d}\; \rm[Fe/H] (r)}{{\rm d}\; \rm{log}(R/R_{\rm{d}})}, 
\end{equation}
where $R_{\rm{d}}$ is the scale length of the progenitor's disc. 
The spatial metallicity distribution of the merger remnants could reveal the initial metallicity gradient of the progenitor galaxy. 
\revise{In fact}, \citet{Fardal2008} showed that disc infalling models \revise{generate} the \revise{differences} of metallicity in the GSS\revise{,} although the initial radial gradient of the progenitor is quite high \revise{($\Delta$\feh$\sim-1.0$, read by eye from figure 2b in \citet{Fardal2008})}. 
\revise{Metallicity gradients in the GSS appear even in models of spherically} symmetric progenitor galaxy \citep{Miki2016}. 
However, \revise{the observed large differences in mean metallicity do not easily develop in these models}. 

To estimate \revise{the} metallicity distribution \revise{in} faint regions\revise{,} such as the broad western structure of the GSS, \revise{we need to reduce} the Poisson noise in the $N$--body \revise{simulations}. 
\revise{For this purpose, we seed the progenitor galaxy with over sixteen million particles in the highest-resolution simulation using GOTHIC (see \autoref{tab:table_resolution}). 
Other parameters for the progenitor model are those of the high-resolution run. }

As \revise{shown in the} top panels of Fig.~\ref{evolution}, most of the disrupted bulge component \revise{appear on the M31 disc}. 
Therefore, we \revise{set the} metallicity gradient only to the disc component of the progenitor galaxy. 
\revise{Initially, we assume} $\Delta$\feh$=-0.5$\revise{, the observed metallicity gradient of} \citep{Koleva2009}. 
\revise{We also set the mean} metallicity to \mfeh$=-0.57$\revise{,} consistent with the mass-metallicity relation of nearby dwarf galaxies \citep{Dekel2003}. 

Fig.~\ref{metal_spatial}b shows the metallicity distribution \revise{in} the disrupted disc of the progenitor galaxy. 
\revise{To} remove extremely faint structures \revise{from} this figure, we set a simple detection limit and \revise{generate} Fig.~\ref{metal_spatial}c. 
\revise{The} total absolute magnitude \revise{of} the GSS \revise{is set to} M$_{\rm{V}}=-14$ \citep{Ibata2001} and the apparent magnitude \revise{limit} in \revise{the} V-band \revise{is} 24.5 (detection limit of INT/WFC). 
\revise{The} initial radial metallicity gradient \revise{yields large difference in the} metallicity distribution \revise{in} the azimuthal direction of the GSS. 
Fig.~\ref{metal_spatial}c suggests that \revise{in the east--west direction,} the stellar population \revise{of the GSS} originates from the \revise{centre} of the initial satellite \revise{progenitor's disc and proceeds toward the outside}. 
Interestingly, \revise{similar} azimuthal \revise{differences} of metallicity \revise{also occurred in} the western shell (see Fig.~\ref{metal_spatial}c). 

\begin{figure}
  \includegraphics[width=\columnwidth]{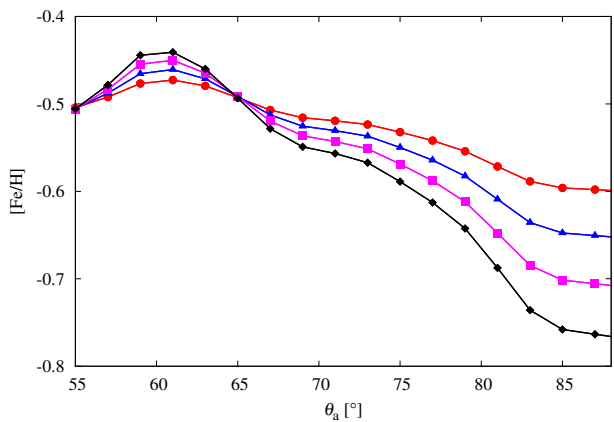}
  \caption{\revise{Mean azimuthal} metallicity distribution \revise{in} the GSS. 
\revise{The inclination of the disc model is THICK7 with $(\phi,\theta)=(-15\degr, 30\degr)$.}
The mean metallicity of the progenitor's disc \mfeh~is set to $-0.5$. 
The initial metallicity gradient of \revise{each line is} $-0.2$ (red line with circles), $-0.3$ (blue line with triangles), $-0.4$ (magenta line with squares) and $-0.5$ (black line with diamonds). 
    \label{metal_different_gradient}
  }
\end{figure}

\revise{\citet{Ibata2007} and \citet{Gilbert2009} observed the mean metallicity in the cocoon region, which is} far from the `core' of the GSS and \revise{contains few simulated particles}. 
Therefore, \revise{we cannot directly compare} the simulation \revise{results with observations in this region}. 
\revise{Instead, we analyse the} azimuthal metallicity distribution \revise{near the} radius of the observed data ($3.5\degr< R<4.5\degr$ from \revise{M31's centre}). 
Fig.~\ref{metal_different_gradient} shows the azimuthal \revise{distribution of the} mean metallicity of the GSS\revise{, where} the mean metallicity of the whole progenitor's disc is $-0.5$\revise{, and} the initial metallicity gradient \revise{is varied} from $-0.5$ to $-0.2$. 
The GSS axis \revise{is} $\theta_a\sim 65\degr$ (see \autoref{tab:table2})\revise{. The} mean metallicity around this axis is relatively \revise{high} and almost \revise{equals} the mean metallicity of the whole progenitor's disc. 
On the other hand, the mean metallicity \revise{in the GSS envelope} ($\theta_a\ga 80\degr$) is relatively \revise{low}. 
We \revise{obtain} strong metallicity \revise{differences} from \revise{the} core \revise{to the} envelope \revise{regions} of the GSS. 

\citet{Fardal2012} observed the mean metallicity in the western shell along the minor axis of the M31 disc and \revise{obtained} \mfeh$=-0.7$. 
Here, we \revise{roughly fit our} simulated mean metallicity to the observed \mfeh \revise{values under two metallicity gradient conditions. For the smaller $\Delta$\feh$=-0.3$ and larger $\Delta$\feh$=-0.5$ metallicity gradients, the best-fitted \mfeh values are $-0.62$ and $-0.57$, respectively}. 

\begin{figure*}
  \includegraphics[width=\textwidth]{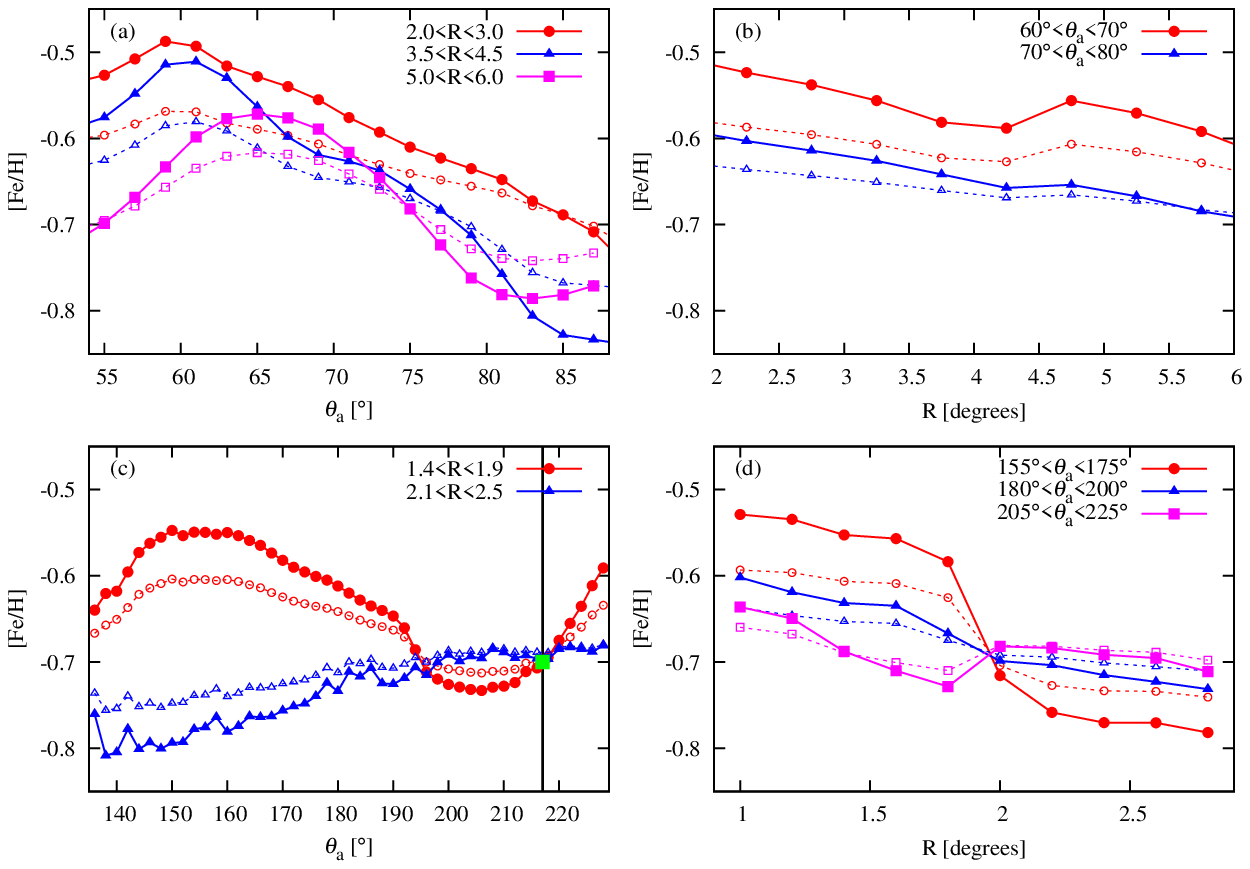}
  \caption{Mean metallicity \revise{distributions in} (a) azimuthal direction of the GSS, (b) radial direction of the GSS, (c) azimuthal direction of the western shell and (d) radial direction of the western shell. 
\revise{The inclination of the disc model is THICK7 with $(\phi,\theta)=(-15\degr, 30\degr)$.}
Solid and dotted lines are the results of ($\Delta$\feh, \mfeh)=($-0.5,-0.57$) and ($\Delta$\feh, \mfeh)=($-0.3,-0.62$), respectively. 
The vertical black solid line \revise{and the green square} in panel (c) \revise{indicate} the minor axis of M31's disc \revise{and} the observed mean metallicity \revise{along} the minor axis \citep{Fardal2012}\revise{, respectively}. 
    \label{metal_gradient}
  }
\end{figure*}

We also \revise{analyse} the azimuthal metallicity \revise{distributions at} similar \revise{radii} of the observed data. 
Fig.~\ref{metal_gradient} \revise{plots} the azimuthal and radial metallicity \revise{distributions} of the GSS and the western shell \revise{for} ($\Delta$\feh, \mfeh)=($-0.5,-0.57$) and \revise{the case of }($\Delta$\feh, \mfeh)=($-0.3,-0.62$). 
\revise{Fig.~\ref{metal_spatial}a shows} the \revise{analysed areas of the mean metallicity distributions in the azimuthal and radial directions (outlined by red solid and blue dashed lines, respectively) of the GSS and} western shell. 
\revise{The} azimuthal mean metallicity \revise{distributions} of the GSS \revise{are plotted in Fig.~\ref{metal_gradient}a}. 
The clearest \revise{metallicity differences appear at} $3.5\degr<R< 4.5\degr$. 
\revise{In the higher metallicity gradient model, the} difference of \revise{metallicities in the} core region ($\theta_a=65\degr$) and envelope region ($\theta_a=85\degr$) \revise{differ by approximately} 0.25 dex. 
\revise{Conversely, as shown in Fig.~\ref{metal_gradient}b}, the radial \revise{metallicity differences in} the GSS \revise{are} small \revise{in narrow azimuthal ranges ($60\degr<\theta_a<70\degr$ and $70\degr<\theta_a<80\degr$). }
\revise{Radial metallicity differences in the GSS were only recently reported by} \citet{Conn2016}. 
Although their data are \revise{azimuthally averaged}, our \revise{results are} qualitatively consistent with \revise{theirs; namely, the metallicity differences are higher} near the root \revise{than in the tail} of the GSS. 
In Fig.~\ref{metal_gradient}c, we show \revise{the} azimuthal metallicity \revise{differences} inside and outside of the western shell. 
\revise{The} metallicity \revise{differs by approximately} 0.2 from south to north \revise{inside the shell, but scarcely differs} outside the shell. 
\revise{Along} the minor axis of M31's disc, the mean \revise{metallicities are similar} inside and outside the western shell. 
\citet{Fardal2012} measured \revise{only the directional} mean \revise{metallicities}, and \revise{their dataset stacks} the \revise{metallicities} of stars inside and outside the western shell. 
Fig.~\ref{metal_gradient}d shows \revise{the} radial metallicity \revise{differences in} the western shell\revise{. The} mean metallicity drastically \revise{decreases} at the edge of the western shell \revise{($155\degr<\theta_a<175\degr$), suggesting largely inhomogeneous} metallicity distribution in the western shell.

\subsection{Progenitor's Bulge and Central MBH}
\label{sec:MBH_bulge}

\revise{We now} discuss the current position of an MBH initially \revise{centred at} the progenitor galaxy of the GSS. 
\revise{According to the} hierarchical structure formation of the universe, MBHs \revise{centralised in} dwarf galaxies should be wandering in the halo of \revise{their} host galaxy. 
The \revise{assumed spherical component} of the GSS \revise{progenitor galaxy} has \revise{an approximate} stellar mass of $10^{8-9}M_{\sun}$. 
\revise{The observed mass of a central MBH correlates with the mass and velocity dispersion of the spherical component \citep{Magorrian1998,Gultekin2009}. }
The velocity dispersion of the bulge $\sigma_{\rm bulge}$ is $\sim50$ km s$^{-1}$ and the \revise{MBH mass} is simply estimated \revise{as} $4\times 10^5 M_{\sun}$ assuming \revise{the} M--$\sigma$ relation \citep{Gultekin2009}. 
\revise{\citet{Baldassare2015} presented a small central MBH, which has a mass of $\sim 5\times 10^5M_{\sun}$, lies on the correlation.} 
\revise{The bulge component of our disc progenitors is somewhat less massive than \revise{in} the spherical model assumed in \citet{Miki2014}. }
Although \revise{not demonstrated in the present} simulation, it is important to notice that the mass of the bulge component can change by \revise{several factors when varying the} bulge--disc ratio and \revise{the} mass-to-luminosity ratio. 

\revise{Our simulations can track the} current position of \revise{the putative} MBH \revise{initially centred in the progenitor}. 
\revise{In fact}, \citet{Miki2014} predicted the current position of the \revise{progenitor} MBH in the halo of M31\revise{, varying the orbits} of the progenitor galaxy of the GSS. 
The radiation spectrum \revise{of} the gas surrounding the MBH \revise{was} estimated \revise{from} the advection-dominated accretion flow (ADAF) model \citep{Kawaguchi2014}\revise{, which reasonably describes this phenomenon}. 
\revise{According to their results, the MBH might} be observed with \revise{currently operating} radio band telescopes such as JVLA and ALMA. 
\revise{They} assumed a spherical progenitor galaxy of the GSS. 
\revise{We emphasise} that the position of the \revise{surviving} bulge component (Fig.~\ref{evolution}d) \revise{approximates} the most suitable orbit \revise{(ID 1)} of \citet{Miki2014} \revise{in} the sky \revise{coordinates}. 
\revise{Despite the different} morphology of \revise{our} progenitor\revise{, its} bulge core component at the best-fitting epoch \revise{approximates} the current position of the MBH predicted by \citet{Miki2014}. 
The bulge component \revise{and} MBH \revise{undergo similar orbital motions under} the gravitational potential of M31\revise{, which essentially controls both orbits}. 
In addition, the progenitor is currently passing through the \revise{apocentre} of its orbital motion with a slow drift velocity\revise{, implying that the current bulge position is relatively long-term. }
Also, the best-fitting epoch of the simulation runs\revise{, as} defined by the edges of the eastern and western shells\revise{, is less variable than when a spherical progenitor is assumed}. 
Therefore, our results are consistent with the current \revise{MBH} position predicted by \citet{Miki2014}. 

One \revise{might expect} that the progenitor's bulge \revise{can trace} the MBH and \revise{appear in} in surface brightness \revise{data} and/or phase-space mass-density \revise{distributions}. 
\revise{For instance, \citet{Davidge2012} located} a north-eastern clump at $(\xi,\eta)=(0.24,0.20)$ \revise{on M31's disc} with an effective radius of $\sim600$~pc. 
\revise{The} position and size of \revise{this} clump are \revise{indicated by the red circle in Fig.~\ref{disk3d}a}. 
\revise{\citet{Davidge2012} estimated the} stellar mass of the clump \revise{as} $3\times 10^8M_{\sun}$\revise{,} consistent with the \revise{mass of the} progenitor's bulge in our models ($\sim 3\times 10^8 M_{\sun}$). 

\begin{figure}
  \includegraphics[width=.99\columnwidth]{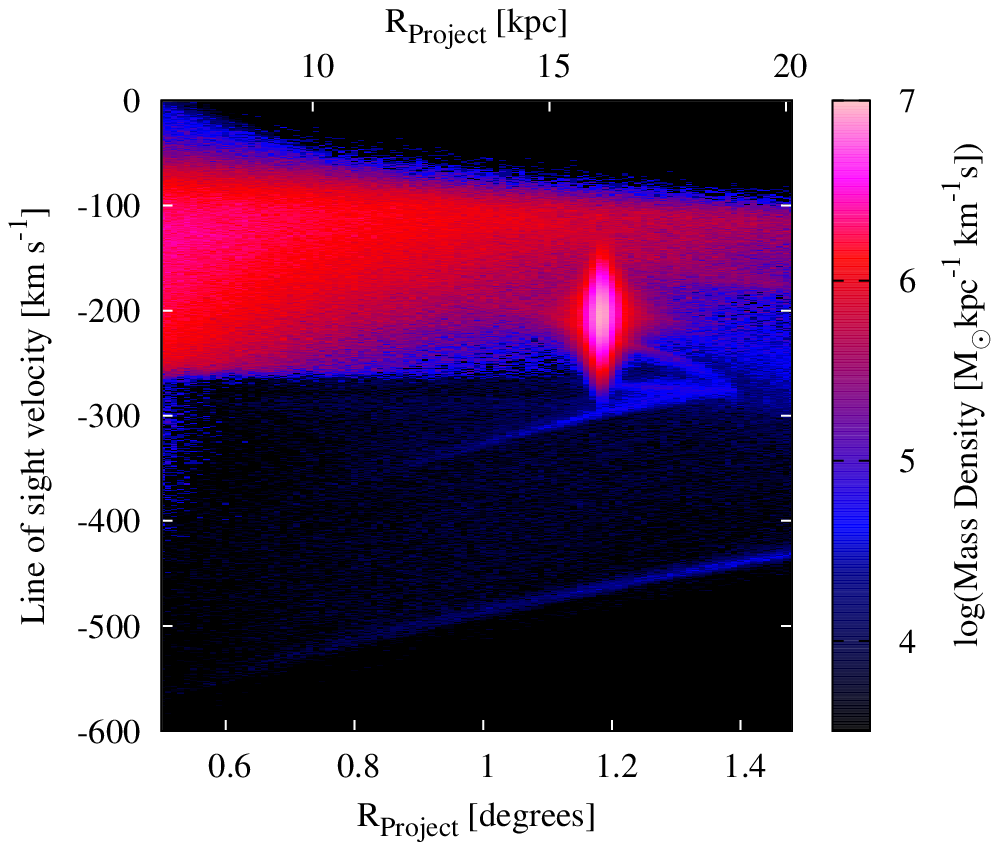}
  \caption{Phase-space mass-density \revise{distributions} (or position-velocity \revise{diagrams}) of the progenitor's bulge and M31 stars. 
\revise{The inclination of the disc model is THICK7 with $(\phi,\theta)=(-15\degr, 30\degr)$.}
    \label{bulge_detection}
  }
\end{figure}

We analyse \revise{the} phase-space distribution \revise{of the progenitor's bulge in the observed frame}. 
\revise{We construct an} M31 model \revise{with} $N$--body particles \revise{in the} MAny-component Galactic Initial-conditions generator (MAGI) (\citeauthor{Miki2017} in prep.), \revise{adopting} the physical quantities \revise{of our present} fixed potential model. 
Fig.~\ref{bulge_detection} shows \revise{the} phase-space mass-density \revise{distributions} of the disrupted progenitor galaxy and M31 stars. 
\revise{This figure is constructed from the same data as Fig.~\ref{highreso}c. }
The \revise{analysed} region is \revise{outlined in} white in Fig.~\ref{disk3d}a. 
The clumpy region at $R\sim15$~kpc \revise{mainly constitutes} the bulge component of the progenitor galaxy. 
Of course, \revise{the total bulge mass and progenitor orbit are uncertain,} but such a bulge remnant should be detected by integral field spectroscopic and/or photometric observations around the predicted position. 
In addition, if high velocity dispersion \revise{(probably induced by the MBH) occurred in the central region,} the bulge component would be easily recognized (see, e.g. \citet{Seth2014}). 

\subsection{Extended stellar shell}
\label{sec:extended_shell}

\revise{Our} unprecedented \revise{highly resolved} simulation \revise{of} the minor merger \revise{enables to predict} a faint but huge stellar structure \revise{outside} the western shell (see Fig.~\ref{metal_spatial}). 
The \revise{outer western shell (hereafter \ews) originates from} outermost region of the initial progenitor galaxy. 
As \revise{shown in} Fig.~\ref{metal_spatial}b and \ref{metal_gradient}c, the metallicity \revise{is lower} in the \revise{{\ews}} than \revise{in the} GSS and \revise{both} shells. 
On the other hand, the \revise{westernmost} side of the GSS is \revise{sourced from the} outermost region of the initial progenitor galaxy and \revise{appears as a broad GSS structure}. 
The \revise{{\ews}} metallicity will also limit the initial metallicity distribution of the satellite progenitor. 
Estimating the surface brightness of the \revise{{\ews}} is important for further observations. 
\revise{A faint extent of metal-poor component has appeared in PAandAS observations (see fig.~2 of \citet{Martin2013}), which may correspond to our simulated faint shell. 
This correspondence requires validation by additional spectroscopic observations of the faint component. }

\begin{figure}
  \includegraphics[width=.99\columnwidth]{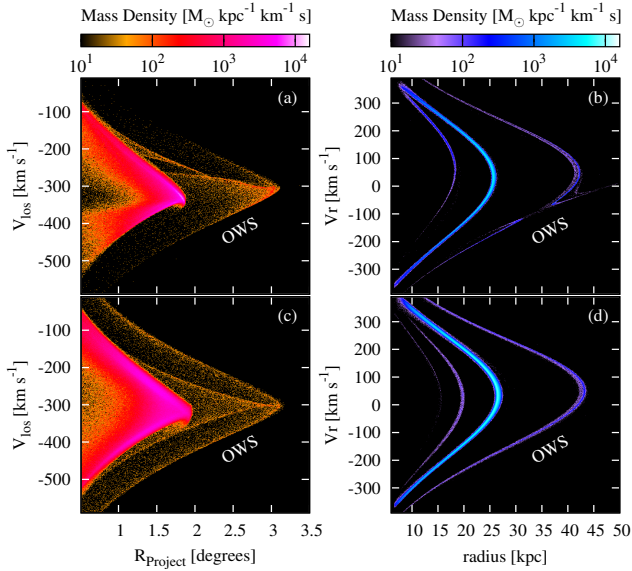}
  \caption{
\revise{
Phase-space mass-density distributions of the simulated western shell and {\ews} assuming the disc (a and b) and Plummer (c and d) progenitors. 
Panels (a) and (c): Line-of-sight velocity distributions. 
Panels (b) and (d): Phase-space distribution centred on M31's centre. 
The inclination of the disc model is THICK7 with $(\phi,\theta)=(-15\degr, 30\degr)$. }
    \label{ews_phase}
  }
\end{figure}

\revise{Fig.~\ref{ews_phase} shows the phase-space mass-density distributions in the western shell and \ews. 
The analysed area is $180\degr<\theta_a<230\degr$ with radii $R > 0.5\degr$. 
This figure is constructed from the data set of the highest-resolution run in the THICK model $(\phi,\theta)=(-15\degr, 30\degr)$. 
To compare the phase-space distributions of the disc and Plummer progenitors, we use data with very similar mass-resolutions in the  two cases. 
Fig.~\ref{ews_phase}a shows the western shell at $R < 2\degr$ and the {\ews} at $R < 3\degr$. 
Both shells are clearly distinguished by their phase-space mass-density distributions (Fig.~\ref{ews_phase}b). 
Panels c and d of Fig.~\ref{ews_phase} reveal a similar structure in a spherical progenitor merger. 
The disc and Plummer models differ in their {\ews} phase-space distributions; specifically, the latter model exhibits a symmetric pattern in $V_{\rm los}$ (Fig.~\ref{ews_phase}c) whereas the former shows an asymmetric pattern (Fig.~\ref{ews_phase}a). }

\begin{figure}
  \includegraphics[width=.99\columnwidth]{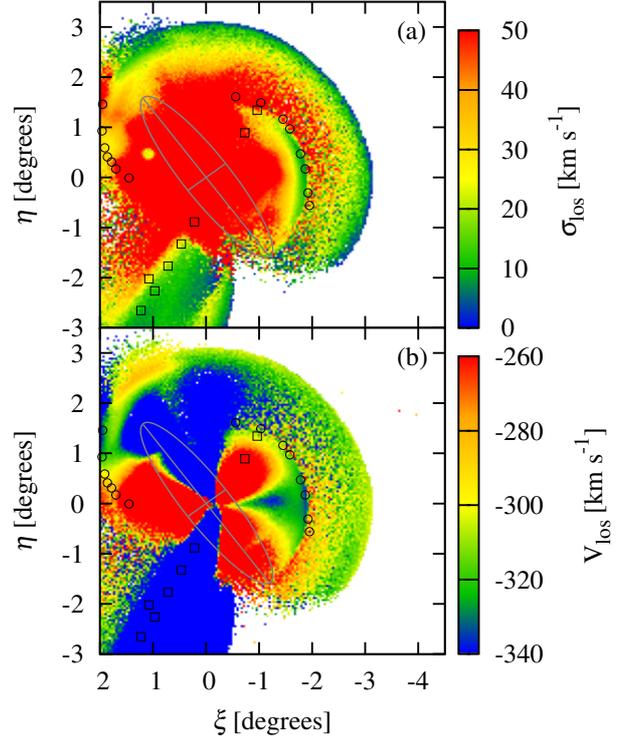}
  \caption{
\revise{
Distribution of line-of-sight velocity dispersion (a) and velocity (b) in the simulated western shell and \ews. 
Symbols and lines are those of Fig.~\ref{best--fitting}a. 
The inclination of the disc model is THICK7 with $(\phi,\theta)=(-15\degr, 30\degr)$. }
    \label{ews_phase2}
  }
\end{figure}

\revise{
Fig.~\ref{ews_phase2} shows the line-of-sight velocity and spatial distribution of the velocity dispersion in the observed frame. 
The line-of-sight velocity dispersion of $\sigma_{\rm los}\simeq0$ reveals clear edges of the western shell and {\ews} (Fig.~\ref{ews_phase2}a). 
The {\ews} particles have experienced two pericentric passages, as particles in the eastern shell. 
Therefore, on the phase-space distribution, the particles in the {\ews} and eastern shell exhibit the same phase. 
}

\revise{At the \ews, the disc and spherical models differ primarily by their distances from us. 
In the Plummer model, the OWS corresponds to a semicircular arc at the bottom-right of Fig.~\ref{disk_plummer3d}f ($-3\degr<\xi<0\degr$ and $-45{\rm kpc}<{\rm d_{M31}}<30{\rm kpc}$). 
In the disc model, most of the stellar components in the OWS spread out only in the foreground of M31 ($-3\degr<\xi<0\degr$ and $-40{\rm kpc}<{\rm d_{M31}}<0{\rm kpc}$). }

\revise{We here summarise} the extended stellar shell. 
\revise{The surface brightness of the extended shell is almost flat and requires a V-band detection limit} 3--4 magnitude deeper \revise{than} the apparent magnitude at the western shell on the minor-axis of M31's disc. 
\revise{The shell is observed in both disc and spherical progenitor models. }
\revise{The morphologies of the progenitor galaxies in the two models differ by their distance between the {\ews} and us}. 
The stars \revise{in} the extended stellar shell have relatively \revise{low metallicities with small} azimuthal \revise{variation,} although the metallicity in the azimuthal direction \revise{varies widely in the simulated western shell}. 
\revise{Future observations} of the extended stellar shell (beyond the currently observed region) \revise{would further constrain the progenitor model}. 

\subsection{Gas distribution}

\revise{The observed} structures \revise{in} the GSS \revise{favour} a rotating disc galaxy as the progenitor. 
\revise{As disc} galaxies \revise{frequently contain} gaseous \revise{components, we can predict that such} components \revise{are stripped and dispersed} in \revise{M31's halo}. 
H\,{\sevensize\bf I} observations around \revise{M31's disc} have \revise{revealed high-velocity} H\,{\sevensize\bf I} clumps that \revise{aligned} the GSS \revise{with an} offset \revise{of} $\sim 15$~kpc \revise{and} a similar line-of-sight velocity \revise{to} the GSS \citep{Westmeier2005,Westmeier2008,Lewis2013}. 
Only recently, another gaseous component \revise{in the GSS was detected in the} absorption spectra of a background Active Galactic Nuclei source \citep{Koch2015}. 
In addition, \revise{the gaseous rings} of M31's disc could \revise{have been formed by} a recent gaseous interaction \citep{Gordon2006}\revise{, reminiscent of the} past gaseous interaction of a gas-rich progenitor. 
\revise{The origin may be revealed by hydrodynamical simulations}.

\section{Summary}
\label{sec:summary}

\revise{Through} detailed simulations of the merger \revise{event} and \revise{comparisons} with observed data, \revise{we have strictly constrained the} physical quantities of M31 and the infalling progenitor\revise{, including the gravitational potential of M31, the progenitor orbit and progenitor mass}. 
\revise{However, the morphology (and dynamics) of the GSS progenitor galaxy have not been detailed here. }
\revise{By simply analysing the} stellar count maps of the GSS in \revise{M31's halo, we characterised the} asymmetric surface brightness profile across the GSS\revise{, which constrains} the morphology of the GSS progenitor\revise{. We also} perform the first large systematic survey of a minor merger with a disc satellite progenitor galaxy. 

We \revise{identified the} parameter space that \revise{properly} reproduces the asymmetric surface brightness of the GSS on the plane $(\phi,\theta)$, which \revise{defines the inclination angle} of the initial disc. 
The \revise{structure was best reconstructed by the} thick disc model ($R_{\rm{d}}=1.1$~kpc, $Z_{\rm{d}}=0.52$~kpc). 
\revise{The} dynamically hot disc model \revise{cannot easily reproduce} the eastern edge of the GSS\revise{,} because the \revise{GSS is broadened by velocity dispersion in this model}. 
On the other hand, \revise{the thin} disc model \revise{struggles} to reproduce the broad western structure\revise{, because it generates an excessive} dynamically cold component. 

\revise{Finally, we summarise} our four \revise{predictions gained from} the \revise{highly resolved} simulations\revise{, which could be verified in future observations}. 
\revise{First}, the progenitor's bulge \revise{currently occupies} the eastern shell and foreground of the disc of M31 and \revise{the two structures are distinguishable in the} phase-space mass-density \revise{distributions}. 
\revise{Second, we expect} clear metallicity differences \revise{in} the merger remnants\revise{, because the} metallicity \revise{clearly differed} in the azimuthal direction at \revise{approximately} $3.5\degr<R<4.5\degr$. 
\revise{The} western shell also \revise{exhibited clear} metallicity \revise{differences} in the azimuthal direction. 
\revise{Third,} an extended stellar shell \revise{should reside} outside the western shell. 
\revise{This} extended shell \revise{should be detectable in} photometric \revise{observations if the} detection limit in \revise{the} V-band \revise{is 3--4 magnitudes deeper than} the apparent magnitude \revise{of} the western shell on the minor-axis of M31's disc. 
Finally, the \revise{western and extended shells contain clearly different stellar populations and observations of their metallicities and/or distances would further constrain the progenitor model}.

\section*{Acknowledgements}

We thank M. J. Irwin for allowing us to use the observed data. 
\revise{We would like to thank the anonymous referee greatly for fruitful and helpful suggestions to improve the paper. }
\revise{We would also like to thank the Center for Computational Sciences, University of Tsukuba, where all of the numerical calculations are carried out on. }
This work was supported by Grant--in--Aid for JSPS Fellows (T.K. 26.348), and the JSPS Grant--in--Aid for Scientific Research (A)(21244013) and (C)(25400222). 
\revise{This work was also supported by the Japan Science and Technology Agency's (JST) CREST program entitled ``Research and Development of Unified Environment on Accelerated Computing and Interconnection for Post-Petascale Era''.}
\revise{T. Kawaguchi was supported in part by a University Research Support Grant from the NAOJ. }



\bibliographystyle{mn2e}
\bibliography{kirihara}

\begin{thebibliography}{60}
\expandafter\ifx\csname natexlab\endcsname\relax\def\natexlab#1{#1}\fi

\bibitem[{{Baldassare} {et~al}\mbox{.}(2015){Baldassare}, {Reines}, {Gallo}, \&
  {Greene}}]{Baldassare2015}
{Baldassare} V.~F., {Reines} A.~E., {Gallo} E., {Greene} J.~E., 2015, \apjl,
  809, L14

\bibitem[{{Barnes} {et~al}\mbox{.}(2006){Barnes}, {Fluke}, {Bourke}, \&
  {Parry}}]{Barnes2006}
{Barnes} D.~G., {Fluke} C.~J., {Bourke} P.~D., {Parry} O.~T., 2006, \pasa, 23,
  82

\bibitem[{{Bender}, {Kormendy} \& {Dehnen}(1996){Bender}, {Kormendy}, \&
  {Dehnen}}]{Bender1996}
{Bender} R., {Kormendy} J., {Dehnen} W., 1996, \apjl, 464, L123

\bibitem[{{Conn} {et~al}\mbox{.}(2016){Conn}, {McMonigal}, {Bate}, {Lewis},
  {Ibata}, {Martin}, {McConnachie}, {Ferguson}, {Irwin}, {Elahi}, {Venn}, \&
  {Mackey}}]{Conn2016}
{Conn} A.~R. {et~al.}, 2016, ArXiv e-prints

\bibitem[{{Davidge}(2012)}]{Davidge2012}
{Davidge} T.~J., 2012, \apjl, 749, L7

\bibitem[{{Dekel} \& {Woo}(2003)}]{Dekel2003}
{Dekel} A., {Woo} J., 2003, \mnras, 344, 1131

\bibitem[{{Fardal} {et~al}\mbox{.}(2008){Fardal}, {Babul}, {Guhathakurta},
  {Gilbert}, \& {Dodge}}]{Fardal2008}
{Fardal} M.~A., {Babul} A., {Guhathakurta} P., {Gilbert} K.~M., {Dodge} C.,
  2008, \apjl, 682, L33

\bibitem[{{Fardal} {et~al}\mbox{.}(2007){Fardal}, {Guhathakurta}, {Babul}, \&
  {McConnachie}}]{Fardal2007}
{Fardal} M.~A., {Guhathakurta} P., {Babul} A., {McConnachie} A.~W., 2007,
  \mnras, 380, 15

\bibitem[{{Fardal} {et~al}\mbox{.}(2012){Fardal}, {Guhathakurta}, {Gilbert},
  {Tollerud}, {Kalirai}, {Tanaka}, {Beaton}, {Chiba}, {Komiyama}, \&
  {Iye}}]{Fardal2012}
{Fardal} M.~A. {et~al.}, 2012, \mnras, 423, 3134

\bibitem[{{Fardal} {et~al}\mbox{.}(2013){Fardal}, {Weinberg}, {Babul}, {Irwin},
  {Guhathakurta}, {Gilbert}, {Ferguson}, {Ibata}, {Lewis}, {Tanvir}, \&
  {Huxor}}]{Fardal2013}
{Fardal} M.~A. {et~al.}, 2013, \mnras, 434, 2779

\bibitem[{{Ferguson} {et~al}\mbox{.}(2004){Ferguson}, {Chapman}, {Ibata},
  {Irwin}, {Lewis}, \& {McConnachie}}]{Ferguson2004}
{Ferguson} A., {Chapman} S., {Ibata} R., {Irwin} M., {Lewis} G., {McConnachie}
  A., 2004, ArXiv Astrophysics e-prints

\bibitem[{{Ferguson} {et~al}\mbox{.}(2002){Ferguson}, {Irwin}, {Ibata},
  {Lewis}, \& {Tanvir}}]{Ferguson2002}
{Ferguson} A.~M.~N., {Irwin} M.~J., {Ibata} R.~A., {Lewis} G.~F., {Tanvir}
  N.~R., 2002, \aj, 124, 1452

\bibitem[{{Font} {et~al}\mbox{.}(2006){Font}, {Johnston}, {Guhathakurta},
  {Majewski}, \& {Rich}}]{Font2006}
{Font} A.~S., {Johnston} K.~V., {Guhathakurta} P., {Majewski} S.~R., {Rich}
  R.~M., 2006, \aj, 131, 1436

\bibitem[{{Gadotti}(2009)}]{Gadotti2009}
{Gadotti} D.~A., 2009, \mnras, 393, 1531

\bibitem[{{Geehan} {et~al}\mbox{.}(2006){Geehan}, {Fardal}, {Babul}, \&
  {Guhathakurta}}]{Geehan2006}
{Geehan} J.~J., {Fardal} M.~A., {Babul} A., {Guhathakurta} P., 2006, \mnras,
  366, 996

\bibitem[{{Gilbert} {et~al}\mbox{.}(2007){Gilbert}, {Fardal}, {Kalirai},
  {Guhathakurta}, {Geha}, {Isler}, {Majewski}, {Ostheimer}, {Patterson},
  {Reitzel}, {Kirby}, \& {Cooper}}]{Gilbert2007}
{Gilbert} K.~M. {et~al.}, 2007, \apj, 668, 245

\bibitem[{{Gilbert} {et~al}\mbox{.}(2009){Gilbert}, {Guhathakurta},
  {Kollipara}, {Beaton}, {Geha}, {Kalirai}, {Kirby}, {Majewski}, \&
  {Patterson}}]{Gilbert2009}
{Gilbert} K.~M. {et~al.}, 2009, \apj, 705, 1275

\bibitem[{{Gordon} {et~al}\mbox{.}(2006){Gordon}, {Bailin}, {Engelbracht},
  {Rieke}, {Misselt}, {Latter}, {Young}, {Ashby}, {Barmby}, {Gibson}, {Hines},
  {Hinz}, {Krause}, {Levine}, {Marleau}, {Noriega-Crespo}, {Stolovy},
  {Thilker}, \& {Werner}}]{Gordon2006}
{Gordon} K.~D. {et~al.}, 2006, \apjl, 638, L87

\bibitem[{{Guhathakurta} {et~al}\mbox{.}(2006){Guhathakurta}, {Rich},
  {Reitzel}, {Cooper}, {Gilbert}, {Majewski}, {Ostheimer}, {Geha}, {Johnston},
  \& {Patterson}}]{Guhathakurta2006}
{Guhathakurta} P. {et~al.}, 2006, \aj, 131, 2497

\bibitem[{{G{\"u}ltekin} {et~al}\mbox{.}(2009){G{\"u}ltekin}, {Richstone},
  {Gebhardt}, {Lauer}, {Tremaine}, {Aller}, {Bender}, {Dressler}, {Faber},
  {Filippenko}, {Green}, {Ho}, {Kormendy}, {Magorrian}, {Pinkney}, \&
  {Siopis}}]{Gultekin2009}
{G{\"u}ltekin} K. {et~al.}, 2009, \apj, 698, 198

\bibitem[{{Hammer} {et~al}\mbox{.}(2010){Hammer}, {Yang}, {Wang}, {Puech},
  {Flores}, \& {Fouquet}}]{Hammer2010}
{Hammer} F., {Yang} Y.~B., {Wang} J.~L., {Puech} M., {Flores} H., {Fouquet} S.,
  2010, \apj, 725, 542

\bibitem[{{Hernquist}(1990)}]{Hernquist1990}
{Hernquist} L., 1990, \apj, 356, 359

\bibitem[{{Ibata} {et~al}\mbox{.}(2004){Ibata}, {Chapman}, {Ferguson}, {Irwin},
  {Lewis}, \& {McConnachie}}]{Ibata2004}
{Ibata} R., {Chapman} S., {Ferguson} A.~M.~N., {Irwin} M., {Lewis} G.,
  {McConnachie} A., 2004, \mnras, 351, 117

\bibitem[{{Ibata} {et~al}\mbox{.}(2001){Ibata}, {Irwin}, {Lewis}, {Ferguson},
  \& {Tanvir}}]{Ibata2001}
{Ibata} R., {Irwin} M., {Lewis} G., {Ferguson} A.~M.~N., {Tanvir} N., 2001,
  \nat, 412, 49

\bibitem[{{Ibata} {et~al}\mbox{.}(2007){Ibata}, {Martin}, {Irwin}, {Chapman},
  {Ferguson}, {Lewis}, \& {McConnachie}}]{Ibata2007}
{Ibata} R., {Martin} N.~F., {Irwin} M., {Chapman} S., {Ferguson} A.~M.~N.,
  {Lewis} G.~F., {McConnachie} A.~W., 2007, \apj, 671, 1591

\bibitem[{{Irwin} {et~al}\mbox{.}(2005){Irwin}, {Ferguson}, {Ibata}, {Lewis},
  \& {Tanvir}}]{Irwin2005}
{Irwin} M.~J., {Ferguson} A.~M.~N., {Ibata} R.~A., {Lewis} G.~F., {Tanvir}
  N.~R., 2005, \apjl, 628, L105

\bibitem[{{Kalirai} {et~al}\mbox{.}(2010){Kalirai}, {Beaton}, {Geha},
  {Gilbert}, {Guhathakurta}, {Kirby}, {Majewski}, {Ostheimer}, {Patterson}, \&
  {Wolf}}]{Kalirai2010}
{Kalirai} J.~S. {et~al.}, 2010, \apj, 711, 671

\bibitem[{{Kalirai} {et~al}\mbox{.}(2006){Kalirai}, {Guhathakurta}, {Gilbert},
  {Reitzel}, {Majewski}, {Rich}, \& {Cooper}}]{Kalirai2006}
{Kalirai} J.~S., {Guhathakurta} P., {Gilbert} K.~M., {Reitzel} D.~B.,
  {Majewski} S.~R., {Rich} R.~M., {Cooper} M.~C., 2006, \apj, 641, 268

\bibitem[{{Kawaguchi} {et~al}\mbox{.}(2014){Kawaguchi}, {Saito}, {Miki}, \&
  {Mori}}]{Kawaguchi2014}
{Kawaguchi} T., {Saito} Y., {Miki} Y., {Mori} M., 2014, \apjl, 789, L13

\bibitem[{{Kirihara}, {Miki} \& {Mori}(2014){Kirihara}, {Miki}, \&
  {Mori}}]{Kirihara2014}
{Kirihara} T., {Miki} Y., {Mori} M., 2014, \pasj, 66, L10

\bibitem[{{Koch} {et~al}\mbox{.}(2015){Koch}, {Danforth}, {Rich}, {Ibata}, \&
  {Keeney}}]{Koch2015}
{Koch} A., {Danforth} C.~W., {Rich} R.~M., {Ibata} R., {Keeney} B.~A., 2015,
  \apj, 807, 153

\bibitem[{{Koch} {et~al}\mbox{.}(2008){Koch}, {Rich}, {Reitzel}, {Martin},
  {Ibata}, {Chapman}, {Majewski}, {Mori}, {Loh}, {Ostheimer}, \&
  {Tanaka}}]{Koch2008}
{Koch} A. {et~al.}, 2008, \apj, 689, 958

\bibitem[{{Koleva} {et~al}\mbox{.}(2009){Koleva}, {Prugniel}, {De Rijcke},
  {Zeilinger}, \& {Michielsen}}]{Koleva2009}
{Koleva} M., {Prugniel} P., {De Rijcke} S., {Zeilinger} W.~W., {Michielsen} D.,
  2009, Astronomische Nachrichten, 330, 960

\bibitem[{{Kuijken} \& {Dubinski}(1994)}]{Kuijken1994}
{Kuijken} K., {Dubinski} J., 1994, \mnras, 269, 13

\bibitem[{{Kuijken} \& {Dubinski}(1995)}]{Kuijken1995}
{Kuijken} K., {Dubinski} J., 1995, \mnras, 277, 1341

\bibitem[{{Lewis} {et~al}\mbox{.}(2013){Lewis}, {Braun}, {McConnachie},
  {Irwin}, {Ibata}, {Chapman}, {Ferguson}, {Martin}, {Fardal}, {Dubinski},
  {Widrow}, {Mackey}, {Babul}, {Tanvir}, \& {Rich}}]{Lewis2013}
{Lewis} G.~F. {et~al.}, 2013, \apj, 763, 4

\bibitem[{{Magorrian} {et~al}\mbox{.}(1998){Magorrian}, {Tremaine},
  {Richstone}, {Bender}, {Bower}, {Dressler}, {Faber}, {Gebhardt}, {Green},
  {Grillmair}, {Kormendy}, \& {Lauer}}]{Magorrian1998}
{Magorrian} J. {et~al.}, 1998, \aj, 115, 2285

\bibitem[{{Magrini} {et~al}\mbox{.}(2016){Magrini}, {Coccato}, {Stanghellini},
  {Casasola}, \& {Galli}}]{Magrini2016}
{Magrini} L., {Coccato} L., {Stanghellini} L., {Casasola} V., {Galli} D., 2016,
  \aap, 588, A91

\bibitem[{{Martin} {et~al}\mbox{.}(2013){Martin}, {Ibata}, {McConnachie},
  {Dougal Mackey}, {Ferguson}, {Irwin}, {Lewis}, \& {Fardal}}]{Martin2013}
{Martin} N.~F., {Ibata} R.~A., {McConnachie} A.~W., {Dougal Mackey} A.,
  {Ferguson} A.~M.~N., {Irwin} M.~J., {Lewis} G.~F., {Fardal} M.~A., 2013,
  \apj, 776, 80

\bibitem[{{McConnachie}(2012)}]{McConnachie2012}
{McConnachie} A.~W., 2012, \aj, 144, 4

\bibitem[{{McConnachie} {et~al}\mbox{.}(2009){McConnachie}, {Irwin}, {Ibata},
  {Dubinski}, {Widrow}, {Martin}, {C{\^o}t{\'e}}, {Dotter}, {Navarro},
  {Ferguson}, {Puzia}, {Lewis}, {Babul}, {Barmby}, {Bienaym{\'e}}, {Chapman},
  {Cockcroft}, {Collins}, {Fardal}, {Harris}, {Huxor}, {Mackey},
  {Pe{\~n}arrubia}, {Rich}, {Richer}, {Siebert}, {Tanvir}, {Valls-Gabaud}, \&
  {Venn}}]{McConnachie2009}
{McConnachie} A.~W. {et~al.}, 2009, \nat, 461, 66

\bibitem[{{McConnachie} {et~al}\mbox{.}(2003){McConnachie}, {Irwin}, {Ibata},
  {Ferguson}, {Lewis}, \& {Tanvir}}]{McConnachie2003}
{McConnachie} A.~W., {Irwin} M.~J., {Ibata} R.~A., {Ferguson} A.~M.~N., {Lewis}
  G.~F., {Tanvir} N., 2003, \mnras, 343, 1335

\bibitem[{{McGaugh} {et~al}\mbox{.}(2000){McGaugh}, {Schombert}, {Bothun}, \&
  {de Blok}}]{McGaugh2000}
{McGaugh} S.~S., {Schombert} J.~M., {Bothun} G.~D., {de Blok} W.~J.~G., 2000,
  \apjl, 533, L99

\bibitem[{{Miki} {et~al}\mbox{.}(2014){Miki}, {Mori}, {Kawaguchi}, \&
  {Saito}}]{Miki2014}
{Miki} Y., {Mori} M., {Kawaguchi} T., {Saito} Y., 2014, \apj, 783, 87

\bibitem[{{Miki}, {Mori} \& {Rich}(2016){Miki}, {Mori}, \& {Rich}}]{Miki2016}
{Miki} Y., {Mori} M., {Rich} R.~M., 2016, \apj, 827, 82

\bibitem[{{Miki} \& {Umemura}(2016)}]{Miki2016b}
{Miki} Y., {Umemura} M., 2016, submitted to New Astronomy

\bibitem[{{Miki} \& {Umemura}(in prep.)}]{Miki2017}
{Miki} Y., {Umemura} M., in prep.

\bibitem[{{Moffett} {et~al}\mbox{.}(2016){Moffett}, {Ingarfield}, {Driver},
  {Robotham}, {Kelvin}, {Lange}, {Me{\v s}tri{\'c}}, {Alpaslan}, {Baldry},
  {Bland-Hawthorn}, {Brough}, {Cluver}, {Davies}, {Holwerda}, {Hopkins},
  {Kafle}, {Kennedy}, {Norberg}, \& {Taylor}}]{Moffett2016}
{Moffett} A.~J. {et~al.}, 2016, \mnras, 457, 1308

\bibitem[{{Mori} \& {Rich}(2008)}]{Mori2008}
{Mori} M., {Rich} R.~M., 2008, \apjl, 674, L77

\bibitem[{{Navarro}, {Frenk} \& {White}(1996){Navarro}, {Frenk}, \&
  {White}}]{Navarro1996}
{Navarro} J.~F., {Frenk} C.~S., {White} S.~D.~M., 1996, \apj, 462, 563

\bibitem[{{Sadoun}, {Mohayaee} \& {Colin}(2014){Sadoun}, {Mohayaee}, \&
  {Colin}}]{Sadoun2014}
{Sadoun} R., {Mohayaee} R., {Colin} J., 2014, \mnras, 442, 160

\bibitem[{{Salmon} \& {Warren}(1994)}]{SalmonWarren1994}
{Salmon} J.~K., {Warren} M.~S., 1994, Journal of Computational Physics, 111,
  136

\bibitem[{{Seth} {et~al}\mbox{.}(2014){Seth}, {van den Bosch}, {Mieske},
  {Baumgardt}, {Brok}, {Strader}, {Neumayer}, {Chilingarian}, {Hilker},
  {McDermid}, {Spitler}, {Brodie}, {Frank}, \& {Walsh}}]{Seth2014}
{Seth} A.~C. {et~al.}, 2014, \nat, 513, 398

\bibitem[{{Spolaor} {et~al}\mbox{.}(2010){Spolaor}, {Hau}, {Forbes}, \&
  {Couch}}]{Spolaor2010}
{Spolaor} M., {Hau} G.~K.~T., {Forbes} D.~A., {Couch} W.~J., 2010, \mnras, 408,
  254

\bibitem[{{Tollerud} {et~al}\mbox{.}(2012){Tollerud}, {Beaton}, {Geha},
  {Bullock}, {Guhathakurta}, {Kalirai}, {Majewski}, {Kirby}, {Gilbert},
  {Yniguez}, {Patterson}, {Ostheimer}, {Cooke}, {Dorman}, {Choudhury}, \&
  {Cooper}}]{Tollerud2012}
{Tollerud} E.~J. {et~al.}, 2012, \apj, 752, 45

\bibitem[{{Toloba} {et~al}\mbox{.}(2011){Toloba}, {Boselli}, {Cenarro},
  {Peletier}, {Gorgas}, {Gil de Paz}, \& {Mu{\~n}oz-Mateos}}]{Toloba2011}
{Toloba} E., {Boselli} A., {Cenarro} A.~J., {Peletier} R.~F., {Gorgas} J., {Gil
  de Paz} A., {Mu{\~n}oz-Mateos} J.~C., 2011, \aap, 526, A114

\bibitem[{{Warren} \& {Salmon}(1993)}]{WarrenSalmon1993}
{Warren} M.~S., {Salmon} J.~K., 1993, in Proceedings of the 1993 ACM/IEEE
  conference on Supercomputing, ACM, pp. 12--21

\bibitem[{{Westmeier}, {Braun} \& {Thilker}(2005){Westmeier}, {Braun}, \&
  {Thilker}}]{Westmeier2005}
{Westmeier} T., {Braun} R., {Thilker} D., 2005, \aap, 436, 101

\bibitem[{{Westmeier}, {Br{\"u}ns} \& {Kerp}(2008){Westmeier}, {Br{\"u}ns}, \&
  {Kerp}}]{Westmeier2008}
{Westmeier} T., {Br{\"u}ns} C., {Kerp} J., 2008, \mnras, 390, 1691

\bibitem[{{Widrow}, {Perrett} \& {Suyu}(2003){Widrow}, {Perrett}, \&
  {Suyu}}]{Widrow2003}
{Widrow} L.~M., {Perrett} K.~M., {Suyu} S.~H., 2003, \apj, 588, 311

\end{thebibliography}








\bsp	
\label{lastpage}
\end{document}